\documentclass[aps,prd,nofootinbib,twocolumn,groupedaddress,showpacs]{revtex4}

\usepackage{graphicx}
\graphicspath{{eps/}}

\begin{document}

\title{Nonlinear radial oscillations of neutron stars}

\author{Michael Gabler$^{1,2}$ and Ulrich Sperhake$^{1,3}$ and Nils Andersson$^{4}$}
\affiliation{${^1}$Theoretisch-Physikalisches Institut,
             Friedrich-Schiller-Universit\"at,
             Max-Wien-Platz 1, 07743 Jena, Germany}
\affiliation{${^2}$ Max-Planck-Institut f\"ur Astrophysik,
             Karl-Schwarzschild-Str. 1, 85741 Garching, Germany}
\affiliation{${^3}$ Theoretical Astrophysics 130-33,
             California Institute of Technology, Pasadena, California 91125, USA}
\affiliation{${^4}$ School of Mathematics, University of Southampton,
             Southampton, SO17 1BJ, United Kingdom}

\date{\today}
\begin{abstract}
The effects of nonlinear oscillations in compact stars are attracting
considerable current interest. In order to study such phenomena
in the framework of fully nonlinear general relativity,
highly accurate numerical studies are required. A numerical scheme
specifically tailored for such a study is based on formulating the
time evolution in terms of deviations from a stationary equilibrium
configuration. Using this technique, we investigate over a wide range
of amplitudes
nonlinear effects in the evolution of radial oscillations of
neutron stars. In particular, we discuss {\em mode coupling} due
to nonlinear interaction, the occurrence of resonance phenomena,
shock formation near the stellar surface as well as the
capacity of nonlinearities to stabilize perturbatively unstable
neutron star models.

\end{abstract}
\maketitle
\section{Introduction}
The observational and theoretical study of stellar oscillations
aimed at gaining insight into stellar structure, an area called
{\em asteroseismology},
has been a rich research
field for a long time. This includes the period-luminosity
relation for Cepheids and the variability of RR Lyrae stars
first discovered by Leavitt and Pickering and Flemming, respectively,
\cite{Leavitt1912, Pickering1901};
see \cite{Alcock1995, Benedict2007, Olech2008} and references
therein for more recent studies. Similarly, much insight into the
structure of the Sun has been obtained via {\em helioseismology}
(see \cite{Birch2008} for a recent review). Oscillations
have also attracted a great deal of attention in the context of
compact objects, neutron stars and black holes; consider, for example,
the stability analysis of neutron stars using radial oscillation
modes by Chandrasekhar \cite{Chandrasekhar1964}.
More recently, interest in neutron star and black-hole
oscillations has focused
on their potential in the context of gravitational wave (GW) physics
\cite{Sathyaprakash2009}.

While most insight into stellar oscillations has been
obtained from linear analysis (see, for example,
\cite{Chanmugam1977,Glass1983,Vaeth1992,Kokkotas2001}),
nonlinear effects are known to play an important role in
the phenomenology of oscillations in various scenarios.
Examples include the nonlinear coupling of modes in the
beat Cepheids (see, for example, \cite{Kollath1998})
and the saturation of $r$-mode oscillations and, thus,
GW generation in rotating
neutron stars first studied by Schenk {\em et al.} \cite{Schenk2002}
and extended in \cite{Arras2003,Brink2004,Brink2004b,Lin2006,Bondarescu2008}.
More generally, perturbative studies have been extended
to include nonlinear
effects up to cubic order terms in the perturbations
\cite{Dziembowski1982,Kumar1989,VanHoolst1996}.
Mode coupling in neutron stars with particular regard to
the generation of GWs has been investigated in the
framework of higher-order perturbation theory in
\cite{Passamonti2006,Passamonti2007}; see also \cite{Brizuela2006,
Brizuela2007, Brizuela2009a, Brizuela2009}
for a gauge-invariant framework to compute
higher-order perturbations including fluid backgrounds.
While the present study focuses on neutron star oscillations, we emphasize
that black-hole oscillations play an equally important role
in GW physics \cite{Berti2006, Ferrari2007}.
At the linearized level, the corresponding solutions are given
in the form of quasinormal modes \cite{Leaver1986,
Kokkotas1999, Berti2007, Nakano2007} which dominate the signal
from the late stages
of a binary black-hole coalescence (see, for example,
\cite{Buonanno2006, Berti2007b, Berti2007, Baker2008a}).
Nonlinearities in black-hole oscillations
have been investigated in the framework
of higher-order perturbation theory as well as numerical
relativity \cite{Nakano2007, Papadopoulos2001}.
A conclusive answer regarding the presence of
nonlinear signatures in ring-down
waveforms resulting from binary-black-hole inspiral and
merger has as yet not been obtained, however, because
of the high numerical accuracy required for such studies
\cite{Berti2007}.

In comparison with black-hole configurations gravity is
about an order of magnitude weaker in spacetimes
containing neutron stars. This has motivated a variety
of hydrodynamic simulations which
implement gravitational effects in the form of some approximation
such as Newtonian theory,
the Cowling approximation or conformal flatness
\cite{Lindblom2000, Lindblom2001, Stergioulas2001, Font2001,
Stergioulas2004, Dimmelmeier2006}.
The present decade has seen
dramatic progress in the numerical solution of the full
Einstein field equations, however, and has resulted
in fully relativistic simulations of single compact stars,
collapse to black holes and neutron star binaries
\cite{Font2002, Baiotti2004,Marronetti2004,Miller2004,Shibata2006,Duez2006,
Anderson2007,Duez2008,Baiotti2008}.
As in the case of black-hole simulations
the available accuracy has not yet reached a level to
facilitate high-precision studies of nonlinear effects, in particular
in the mildly nonlinear regime.

In consequence, there currently exists a
gap in the literature studying in the fully nonlinear
general relativistic framework with high precision
mildly and moderately nonlinear effects in oscillations of compact stars.
The main purpose of the present paper is to fill this gap
in the case of the simplest type of
stellar pulsations, radial oscillations of
spherically symmetric polytropic stellar models. We are aware
that more realistic simulations of neutron stars require the
inclusion of a multitude of microphysical effects such as
more realistic equations of state and magnetic fields.
For this reason, our study is intended in the first place to provide
a general taxonomy of nonlinear features that will be encountered
in the context of stellar pulsations. The
results also provide valuable tools for calibrating the accuracy
of general relativistic three-dimensional hydro codes.

Following Refs.~\cite{Sperhake2001paper, Papadopoulos2001},
we achieve the necessary accuracy by formulating the problem
in terms of deviations from an equilibrium background model,
as is commonly done in traditional perturbation theory. In contrast
to that approximation, however, we keep all terms of higher order
in the deviations and thus arrive at a system of equations
equivalent to the original nonlinear system. The main
advantage of this approach is the elimination from the
equations of all terms exclusively containing background quantities
and, thus, the discretization error associated with these terms.
The key improvement over the toy problem studied in
\cite{Sperhake2001paper} is the inclusion of the entire
star including the important outer layers near the stellar
surface.

This paper is organized in five sections. We set up the numerical
framework in Sec.~\ref{sec: numerics}.
In Sec.~\ref{sec:modecoupling}
we study in detail the {\em coupling} of eigenmodes due to nonlinear
effects, including a more detailed discussion of
saturation and resonance effects. Section \ref{sec: further_effects}
investigates two further nonlinear effects not directly concerning
{\em mode coupling}, the stabilization of linearly unstable stars
and the formation of discontinuities near the stellar surface.
Finally, we summarize our findings in Sec.~\ref{sec:summary}.
Throughout this work we adopt units corresponding to $c=G=1$, where
$c$ is the speed of light and $G$ is the gravitational constant.

\section{Numerical framework}
\label{sec: numerics}

\subsection{Evolution equations}\label{sec: evol_eqs}

The starting point for our description of a dynamic, spherically symmetric
neutron star is the formulation developed
by May and White \cite{MayWhite1966, MisnerSharp1964}.
Specifically, we write the metric in the form
\begin{equation}
  ds^2 = -\lambda^2 dt^2 + \mu^2 dx^2 + r^2 (d\theta^2 + \sin^2 \theta
         d\phi^2),
\end{equation}
where $\lambda$, $\mu$ and areal radius $r$
are functions of time $t$ and radial position
$x$. Unless stated otherwise,
the coordinate $x$ is initialized by the areal radius
of the background configuration (cf.~Sec.~\ref{sec: deviations} below)
and serves as a Lagrangian coordinate following the motion of the
fluid elements during the time evolution.
The neutron star is modeled as a single component perfect fluid
at zero temperature. The corresponding energy momentum tensor can be
expressed in the form $T^{\mu\nu} = (\rho+P) u^\mu u^\nu + P g^{\mu\nu}$,
where $\rho$ and $P$ are energy density and pressure. The four velocity
$u^\mu$ obeys the normalization condition
$u^\mu u_\mu = -1$.  Because the radial coordinate
$x$ is comoving with the fluid elements we have the simple
expression $u^{\mu} = [\lambda^{-1},0,0,0]$.
We further assume rest mass conservation and neglect all heat transfer
other than that due to the motion of the fluid.  In particular, this excludes
heat transfer by neutrinos or radiation as well as pair production and
interaction with external fields.  The equation of state (EOS) is modeled
by a polytropic law $P=K \rho^\gamma$, where $K$ is the polytropic constant
and $\gamma$ the polytropic index.

In order to achieve high accuracy near the boundaries, we find it
important to employ variables with at most linear asymptotic
behavior at both the center and the stellar surface \footnote{See
Sec.~V B.1 in \cite{Seidel1990} for a discussion of the difficulties
arising from higher-order falloff of variables near the boundaries.}.
We therefore describe the stellar model
in terms of a rescaled mass \footnote{A straightforward calculation
shows that $m\equiv r^2N$ evaluated at the surface of the star
is the mass parameter of the exterior vacuum metric in
Schwarzschild coordinates.}
function $N$ and a function
$\sigma$ which are defined by
\begin{eqnarray}
 N &\equiv& 
\frac{1}{2r}\left(
      1-\frac{1}{\mu^2}\right)~\text{,} \\
 \sigma &=& \frac{P}{\rho}~\text{.} \label{eq: def_sigma}
\end{eqnarray}
The set of variables is completed by the metric component $\lambda$
and an auxiliary velocity function $w\equiv r_{,t}/\lambda$ introduced to
obtain a system of partial differential equations of first order
in space and time.

The Einstein equations $ G_{\mu\nu}= 8 \pi  T_{\mu\nu}$
and the conservation of energy and momentum $\nabla_\mu  T^{\mu\nu}=0$
then result in the following equations:
\begin{eqnarray}
  \sigma_{,x} &=& -\frac{\lambda_{,x}}{\lambda}
       \left( 1 + \sigma \right) \left( 1-\frac{\sigma}{C^2}
       \right), \label{MaWdsigmadx}\\
N_{,x} &=& r_{,x} \left( 4 \pi \rho - 2 \frac{N}{r}
       \right), \label{MaWdNdx} \\
N_{,t} &=& - w \lambda \left( 4 \pi P + 2
       \frac{N}{r} \right), \\
r_{,t} &=& \lambda w,\\
w_{,t} &=& (1+w^2 - 2 N r) \frac{\lambda_{,x}}{
     r_{,x}} - \lambda ( N  +4\pi P r) \label{wt},
     \label{eq: wt} \\
\rho_{,t} &=& - \lambda (\rho + P) \left(
      \frac{w_{,x}}{r_{,x}}+\frac{2w}{r}\right)\label{eq:constraint}~\text{.}
\end{eqnarray}
In these equations, commas are used to represent partial derivatives and $C^2=\partial p/\partial\rho$ 
is the sound speed.
Only five of these equations are independent and we choose to use
Eq.~(\ref{eq:constraint}) in the determination of the eigenmodes below
but not in the nonlinear time evolution.

In order to determine boundary conditions, we first consider the origin.
Because the physical system is spherically symmetric by construction, all
functions that are odd in the radius necessarily vanish at the origin, that is
$r(0)=0$, $N(0)=0$ and $w(0)=0$.
The surface $x_{\rm S}$
of the star is defined as the radius where the pressure
vanishes. We further define our time coordinate such that the interior
solution matches to the Schwarzschild metric in the exterior.
We thus obtain outer boundary conditions
$P(x_{\rm S})=0$, corresponding to $\sigma(x_{\rm S})=0$, and
$\lambda(x_{\rm S}) = \sqrt{1-2 N(x_{\rm S}) r(x_{\rm S})}$.

\subsection{Deviations from an equilibrium configuration}
\label{sec: deviations}

The purpose of this study is to analyze nonlinear effects in the
time evolution of
radial oscillations of a neutron star model with particular regard to
the mildly nonlinear regime. Such an investigation requires
numerical simulations of high accuracy.
We achieve this by formulating the numerical evolution
in terms of deviations from a stationary {\em background}
configuration. In such a formulation all
terms involving exclusively background quantities and, thus,
the discretization errors associated with these terms
drop out of the equations. Because we are mostly interested in
scenarios where the deviations are small in comparison with the
background terms, the elimination of these error terms
leads to a significant increase in numerical accuracy.

This decomposition
requires us to choose a convenient
background. For the study at hand, the obvious choice
is a spherically symmetric, static neutron star,
as described by Tolman, Oppenheimer and Volkoff (TOV)
\cite{Tolman1939, OpVol1939}.
The system of equations governing the TOV equilibrium configuration
is given by the time independent limit of the system of Eqs.
(\ref{MaWdsigmadx})-(\ref{eq: wt}),
\begin{eqnarray}
  \bar{\sigma}_{,x} &=& -\frac{\bar{\lambda}_{,x}}{\bar{\lambda}}
       \left( 1 + \bar{\sigma} \right) \left( 1-\frac{\bar{\sigma}}{\bar{C}^2}
       \right) \label{eq: TOV sigma_x} \\
\bar{N}_{,x} &=& \bar{r}_{,x} \left( 4 \pi \bar{\rho} - 2 \frac{\bar{N}}{\bar{r}}
       \right) \\
\frac{\bar{\lambda}_{,x}}{\bar{r}_{,x}} &=& \frac{\bar{\lambda}}
       {1-2\bar{N}\bar{r}} (\bar{N}+4\pi \bar{P}\bar{r}),
       \label{eq: TOV lambda_x}
\end{eqnarray}
where we have introduced an overbar to distinguish background variables
from their time dependent counterparts. Note that $\bar{r}_{,x}=1$
if we initialize $x$ by the areal radius of the background model.
Next, we formally introduce deviations from this equilibrium by
\begin{equation}
 f(x,t) = \bar{f}(x) + \Delta f(x,t).
 \label{eq: decomposition}
\end{equation}
Here and in the following, the
function $f$ stands for either of the variables
$\lbrace \lambda , N , \sigma, \rho, P, r \rbrace$.
Following customary notation in the literature, we denote the
radial displacement by $\xi \equiv \Delta r$.

We can now insert Eq.~(\ref{eq: decomposition}) into the set of evolution
Eqs. (\ref{MaWdsigmadx})-(\ref{eq: wt}). Of particular
significance in the resulting expressions are the terms containing
exclusively background quantities and no deviations. A straightforward
calculation demonstrates that all these terms drop out of the equations
because, by construction, the background variables $\bar{f}$
obey the stationary
limit of Eqs.~(\ref{MaWdsigmadx})-(\ref{eq: wt}), that is, the TOV equations.
We are left exclusively with terms of linear or higher order
in the deviations $\Delta f$ and the evolution equations are given by
\begin{eqnarray}
   \Delta\sigma_{,x} &=&- \frac{\bar{\sigma}_{,x}(\Delta\lambda C^2
        +\bar{\lambda}\Delta C^2)}{\lambda C^2}
        - \frac{\Delta \lambda_{,x} (1+\sigma)
        (C^2 -\sigma)}{\lambda C^2} \nonumber\\
     && - \frac{\bar{\lambda}_{,x}\left[\Delta\sigma (C^2 -\sigma)
        + (1+\bar{\sigma})(\Delta C^2 - \Delta\sigma)\right]}{\lambda C^2}
        \label{nonlinear-sigma} \\
\Delta N_{,x} &=& \bar{r}_{,x} 4\pi\Delta\rho +\xi_{,x} 4 \pi \rho-\nonumber\\
     && \frac{ \bar{N}_{,x}\xi +\xi_x 2 N+\bar{r}_{,x}
        ( 2\Delta N-4\pi\bar{\rho}\xi)}
        {r} \label{nonlinear-Nx}\\
\Delta N_{,t} &=& - w\frac{\lambda }{r}
        \left(4\pi r P + 2 N \right) \label{nonlinear-N}\\
\xi_{,t}&=&  \lambda w \\
w_{,t}&=& \left(1-2\bar{N}\bar{r}\right) \frac{ \Delta\lambda_{,x}} {r_{,x}}
        + (w^2- 2\Delta N r-2 \bar{N} \xi )
        \frac{\lambda_{,x}}{r_{,x}} \nonumber\\
     && -\frac{\xi_x}{r_{,x}} \left( N \lambda
        + 4\pi P r \lambda \right)
        -\frac{\bar{r}_{,x}}{r_{,x}} \left(\Delta N \lambda
        -\Delta \lambda \bar{N} \right) \nonumber\\
    &&  - \frac{\bar{r}_{,x}}{r_{,x}}4\pi\left[\bar{P}  (\bar{r} \Delta\lambda
        +\xi \lambda)+ \Delta P r \lambda\right]
        \label{nonlinear-w}
\end{eqnarray}

The boundary conditions for the deviations follow directly from
those obtained for the full variables $f$ in Sec.~\ref{sec: evol_eqs}.
They are
given by $\xi(0)=0$, $\Delta N(0)=0$, $w(0)=0$ as well as
$\Delta \sigma(x_{\rm s})=0$ and
$\Delta \lambda (\lambda + \Delta \lambda) = -2 (\xi N + \bar{r}\Delta N)$ .

\subsection{Eigenmodes}
\label{sec: eigenmodes}
In the limit of infinitesimally small deviations $\Delta f$ from the
equilibrium solution $\bar{f}$, the time evolution of the star is governed
by the linearized version of the system (\ref{nonlinear-sigma}) -
(\ref{nonlinear-w}). The solutions to the linearized perturbation
equations are given by an infinite set of oscillation modes
with characteristic frequencies. This set of modes is a vital
asset when interpreting the nonlinear time evolution.
For each neutron star model, we therefore calculate the first
10 eigenmode profiles and their associated frequencies.

The linearized evolution system is most conveniently written in terms
of the rescaled radial displacement
$\zeta = \bar{r}^2\xi/\bar{\lambda}$~(cf.~chapter 26 in \cite{Misner1973})
and reduces to the single partial
differential equation
\begin{equation}
 W \zeta_{,t,t} = \frac{1}{\bar{r}_{,x}} \left( \frac{\Pi}{\bar{r}_{,x}}
     \zeta_{,x}\right)_{,x} + Q \zeta~\text{.} \label{eigenmode}
\end{equation}
Here, the variables $W$, $\Pi$ and $Q$ depend only on the background
quantities. Exact expressions for these auxiliary
functions are given in Eqs.~(\ref{Pi})-(\ref{eq: Q}).

Separation of variables straightforwardly implies a harmonic time
dependence, i.~e.~$\zeta \sim e^{i\omega t}$, where $\omega$ is a
frequency yet to be determined.  The resulting equation represents a
singular Sturm-Liouville problem, and the solutions $\zeta_i(x)$ of
this eigenvalue problem form a complete orthonormal set. It is possible,
therefore, to expand the radial displacement function $\zeta (t,x)$
in a series
\begin{equation}
 \zeta(t,x) = \sum_i A_i(t) \zeta_i(x) \label{eq: zeta_expansion},
\end{equation}
where the eigenmode coefficients $A_i(t)$ are given by the inner product
\begin{equation}
  A_i(t)= \langle \zeta, \zeta_i \rangle \equiv
      \int_0^{x_{\rm S}} W \zeta (t,x)
      \zeta_i(x)  ~dx ~\text{.} \label{eq: Ai}
\end{equation}
The $A_i(t)$ thus represent a measure of how strongly an eigenmode $i$
contributes at a given time $t$.

\subsection{Numerical methods}
The equations for the TOV background (\ref{eq: TOV sigma_x})-(\ref{eq:
TOV lambda_x}) as well as the Sturm-Liouville problem (\ref{eigenmode})
represent two-point boundary-value problems which can be solved
by means of a relaxation method \cite{Press1992}.  We discretize
the evolution system (\ref{nonlinear-sigma})-(\ref{nonlinear-w})
using the implicit, second-order accurate
Crank-Nicholson scheme. From a numerical point
of view, this implicit evolution scheme turns out to be conceptually
identical to the two-point boundary-value problems of the background and
eigenmode calculation. We therefore use the same relaxation algorithm
for all of these calculations.

Finally, we compute the integrals
appearing in the calculation of the inner products
using a standard fourth-order accurate Simpson's Rule algorithm.

\subsection{Code tests}
\label{sec: code_tests}
\begin{figure}[b!]\vspace{-2mm}
\centering
 \includegraphics[width=0.4\textwidth]{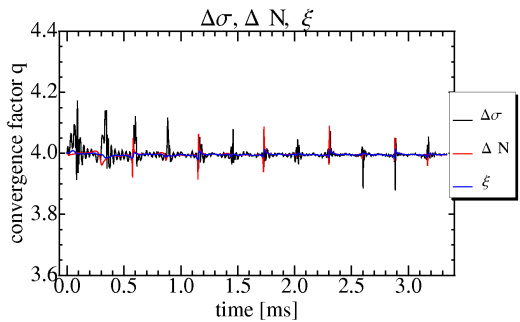}
\caption{The convergence factor $q$ of  $\Delta\sigma$, $\Delta N$ and
         $\xi$ as functions of time. The obtained results are consistent
         with second-order convergence corresponding to $q=4$.
        }
\label{fig:conv}
\end{figure}
The code has been tested in several ways. First, we have performed
a convergence analysis for a polytropic model with $\gamma=2$ and
$K=150~{\rm km}^2$. We use 801, 1601 and 3201
grid points and obtain second-order convergence for the background
model, the calculation of the eigenmode profiles and the time evolution
including the eigenmode coefficients. In Fig.~\ref{fig:conv}, we show
the convergence factors of the $\ell_2$ norms of the
evolution variables $\Delta\sigma$, $\Delta N$ and $\xi$ for a
simulation of a perturbation given initially in the form of the first
eigenmode with an amplitude of $10$~m. The convergence factor
is defined as
\begin{equation}
  q \equiv \frac{\ell_2[f_{\rm c}] - \ell_2[f_{\rm m}]}
        {\ell_2 [f_{\rm m}] - \ell_2 [f_{\rm f}]},
\end{equation}
where the subscripts $c$, $m$ and $f$ stand for coarse, medium and fine
resolution. For a second-order scheme and the above-mentioned
grid setups we expect $q=4$.
The minor deviations from this value
are due to zero crossings of the variables involved.

For the second test, we use the constraint Eq. (\ref{eq:constraint})
which vanishes in the continuum limit. In order to quantify the
constraint violation due to the discretization error,
we normalize the numerical constraint by the
sum of the $\ell_2$ norms of the individual terms.
We find the resulting normalized constraint violations to be
of the order of $10^{-6}$ or less.

Third, we test the eigenmode calculation by checking the completeness
of the eigenmode spectrum. This is done by calculating the weighted sum
 $\sum_i A_i^2 / \langle \zeta, \zeta \rangle$, which has to be
unity due to the completeness of the basis.
In our simulations we include the first 10 terms in this infinite series
and obtain a maximal deviation from the expected value of $1$ by
about $10^{-4}$. As a by-product, this result demonstrates that
the first 10 eigenmodes capture most of the dynamics of the system.

Finally, we compare the eigenmode frequencies with results available in
the literature and observe excellent agreement;
the maximal deviation in the first three eigenmodes
is about 1\% from results reported in \cite{Kokkotas2001} and
less than 1\% from those of \cite{Sperhake2001}.

Using the setup described in this section, we obtain
highly accurate time evolutions. In fact, we observe that the overall
error is dominated by the calculation of the inner products
in Eq.~(\ref{eq: Ai}) via the Simpson algorithm,
rather than the time evolution of the
grid variables. We believe this to be an artifact of the uncertainties
in the eigenmode profiles themselves, the relative error being of the
order of $10^{-5}$.
Motivated by this assumption, we managed to further improve the accuracy
using the fact that
our time evolutions are typically dominated by
one particular eigenmode. First, we identify this mode from
the initial data construction. In the course of the evolution, we
calculate the eigenmode coefficient associated with this mode.
Before calculating the other eigenmode coefficients, however, we
subtract the contribution of the dominating mode from the grid
variables. By virtue of the orthogonality of the eigenmodes,
this subtraction does not affect the other coefficients in the
continuum limit. For finite numerical resolutions, however, the
eigenmodes are not perfectly orthogonal and we
avoid contamination of the overlap integrals due to the dominant
mode.

Taking into account all numerical effects, we arrive at the following
estimates for the uncertainties. All simulations discussed in
the next section start
from initial data consisting of a single mode. This mode is found to
dominate the ensuing evolution and we measure its amplitude with a
relative error of $10^{-5}$. All other modes are absent in the initial
data, but are {\em excited} due to nonlinear effects. The accuracy of their
measurement depends on their amplitude. For those modes used in our
analysis we obtain relative uncertainties ranging from $10^{-2}$
for weak {\em excitation} to $10^{-5}$ for strong {\em excitation}.


\section{Mode coupling}\label{sec:modecoupling}

Before analyzing in detail the {\em coupling} of eigenmodes in our
simulations, we need to address a conceptual difficulty arising from
the nonlinear nature of hydrodynamics.
Eigenmodes are, by construction, a feature of a linear theory and
our calculation of the eigenmode spectrum in Sec.~\ref{sec: eigenmodes}
required us to specify a background configuration.
Given a dynamic
system evolved with the fully nonlinear theory, we analyze the deviations
from that background configuration by projecting them onto the eigenmodes
associated with that same background. The problem is that there exists
no unique way of decomposing such a dynamic
system into a background plus perturbations.
Different background configurations imply different
eigenmode spectra. In consequence of this ambiguity, we can interprete
nonzero mode coefficients for modes not present in the initial data in
two ways; (i) excitation due to nonlinear effects and (ii) a change
in the background configuration and its eigenmode spectrum due
to the finite amplitude of the initial perturbation. Both interpretations
reflect the nonlinear nature of the theory and the nonuniqueness of
choosing a background makes it impossible to distinguish in a well-defined
manner between those interpretations. In the remainder of this work we
will use the terminology {\em mode coupling} and {\em excitation}
but, as a reminder of the
ambiguities in the interpretation, we will put the words in italics.
Note, however, that once a particular choice of the background
has been specified, the ensuing projection of deviations from that
background onto the corresponding eigenmode spectrum is uniquely
defined. In that sense, the following analysis is self-consistent
and any study agreeing upon the same decomposition can be compared
straightforwardly with our results.

Specifically,
our analysis is based on the following construction of a background
model plus nonlinear deviations.
We consider either of the two stellar models labeled 1 and 2
in Table \ref{tab:models}.
Numerically, we prescribe the initial perturbation of this background
model by displacing the fluid elements
from their equilibrium positions according to the profile of
the displacement $\xi$ corresponding to one single eigenmode.
From this displacement, we straightforwardly calculate the profiles
for the remaining evolution variables $\Delta \lambda$, $\Delta \sigma$
and $\Delta N$ according to Eqs.~(\ref{nonlinear-w}),
(\ref{nonlinear-sigma}) and (\ref{nonlinear-Nx}) in that order.
Finally, we set the radial velocity $w$ to zero.
In the remainder of this work, we always label the single mode
present in the initial data as $j$ and denote its amplitude in the
form of its surface displacement $\xi_j \equiv \xi(x_{\rm S})$
Similarly, we
use the label $i$ to identify all modes present in the
time evolution.
\begin{table}[t]
  \begin{center}
  \begin{tabular}{l|ccccc}
    \hline \hline
    Model  &  $\quad\gamma\quad$  &  $K$  &  $\rho_{\rm c}~[{\rm g/cm}^3]$  &  $M~[M_{\odot}]$  &  $R~[{\rm km}]$  \\
    \hline
    1  &  2     &  $150~{\rm km}^2$      &  $2.022 \times 10^{15}$  &  1.555  &  10.82  \\
    2  &  2.25  &  $700~{\rm km}^{2.5}$  &  $3.600 \times 10^{15}$  &  1.584  &  8.414  \\
    3  &  2     &  $150~{\rm km}^2$      &  $3.774 \times 10^{15}$  &  1.655  &  9.222  \\
    \hline \hline
  \end{tabular}
  \end{center}
  \caption{Physical parameters for the three stellar models studied in this
           work.
          }
  \label{tab:models}
\end{table}
\begin{figure*}[t!]
\centering
 \includegraphics[width=.325\textwidth]{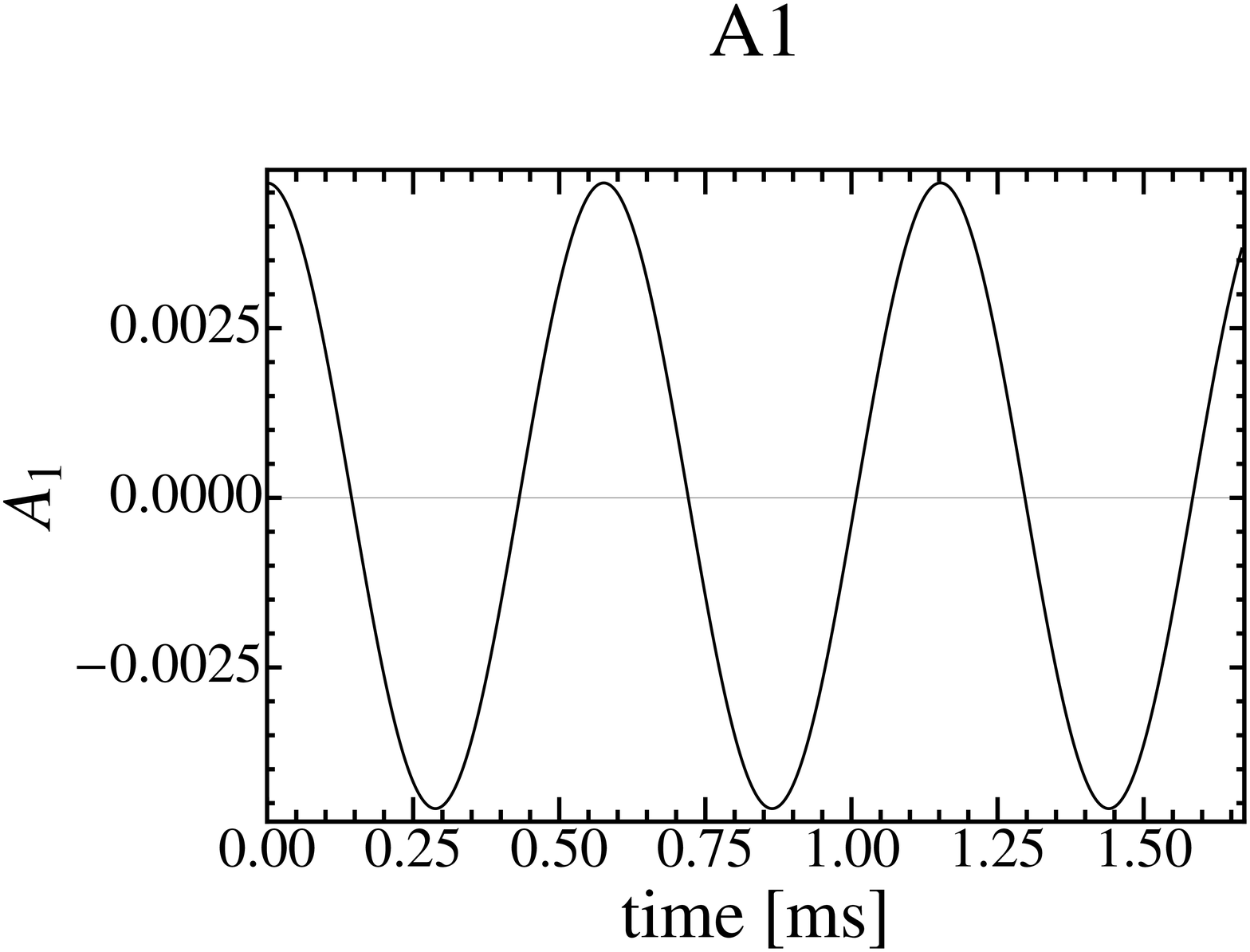}
 \includegraphics[width=.325\textwidth]{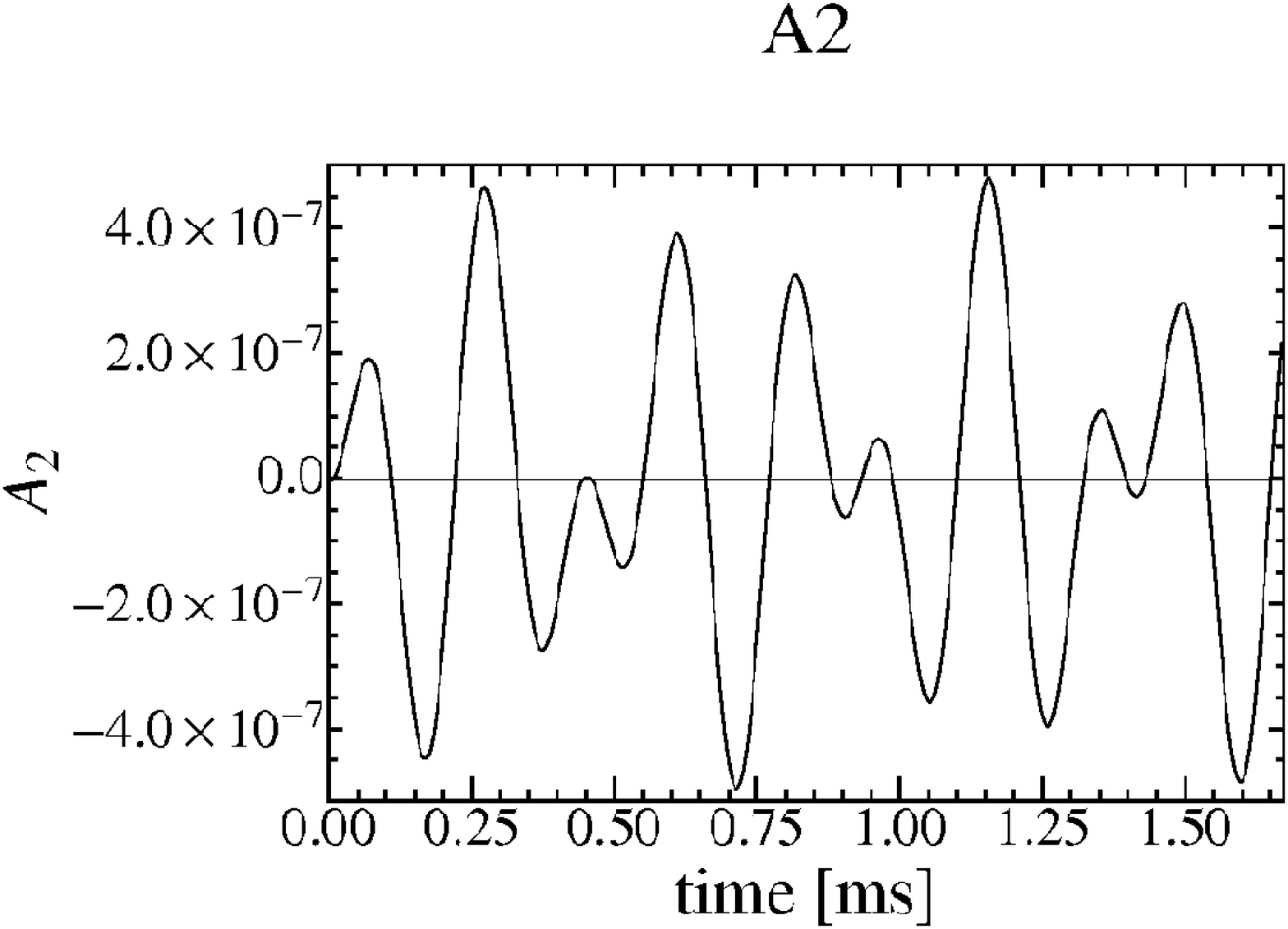}\\

 \includegraphics[width=.325\textwidth]{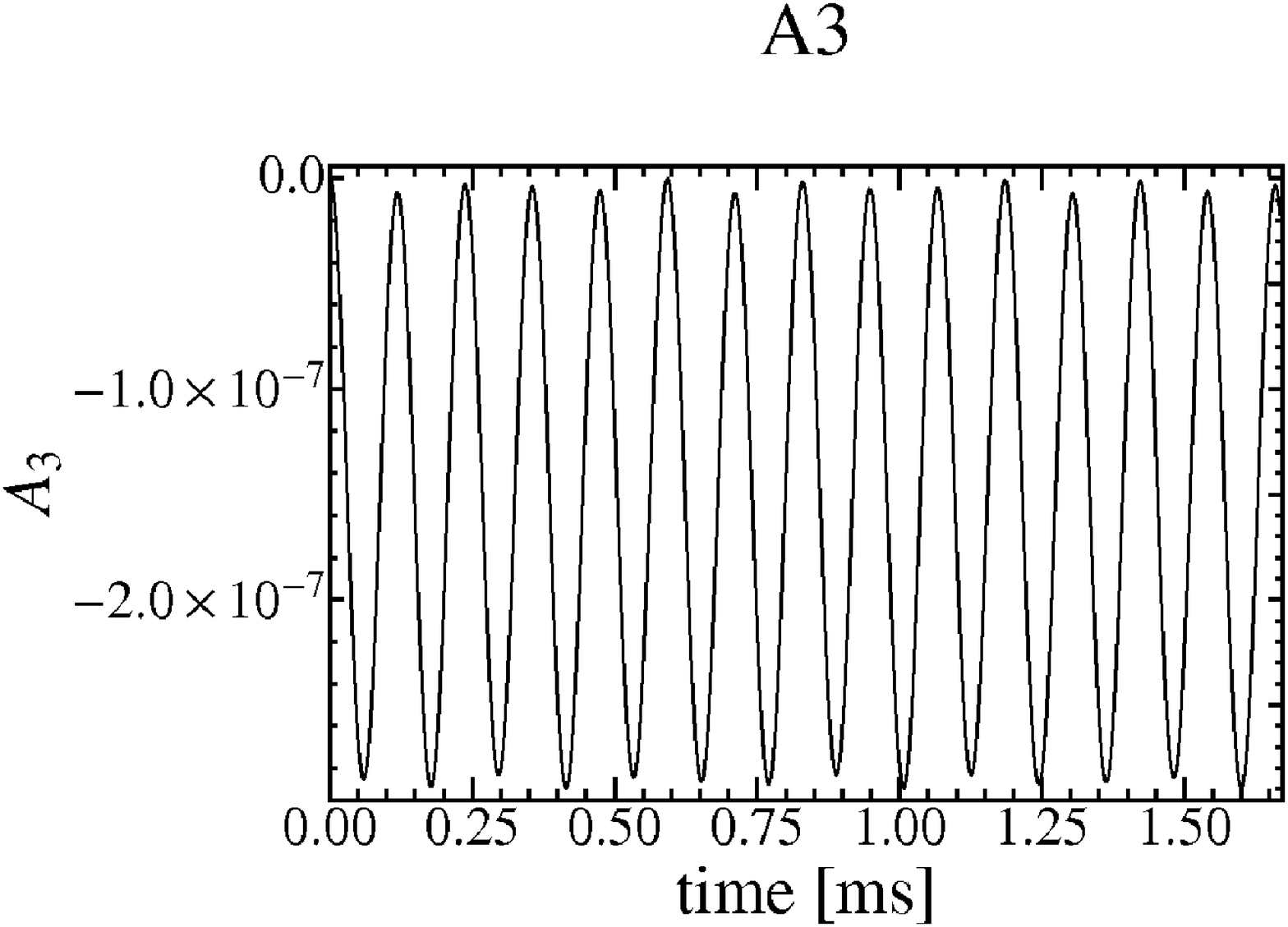}
 \includegraphics[width=.325\textwidth]{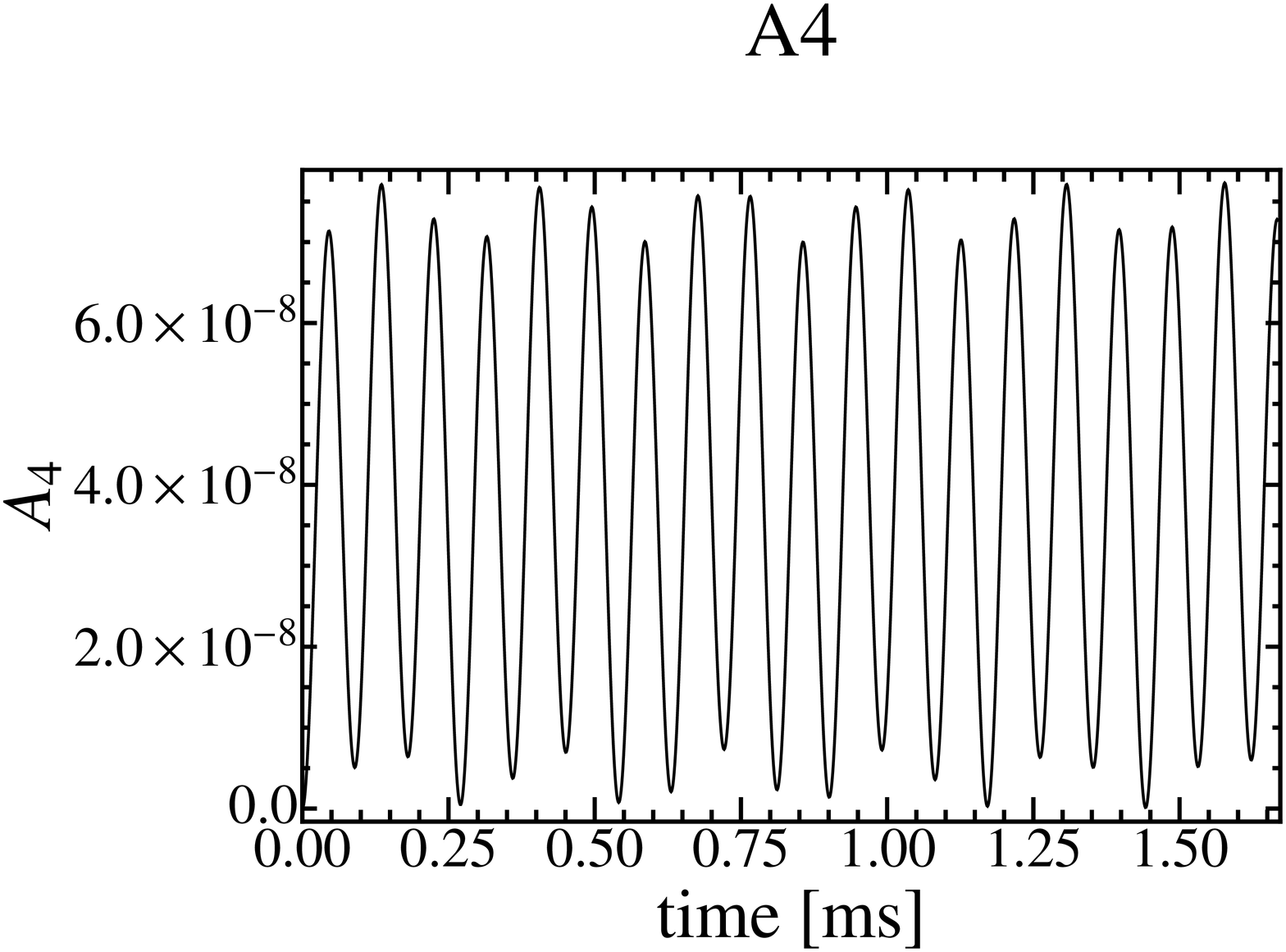}
\caption{The evolution of the first four eigenmode coefficients $A_i$
         obtained for model 1 and an initial displacement corresponding to
         the fundamental mode with a surface amplitude of
         $\xi_{1}=10~{\rm m}$
        }
\label{fig: Ai_case1}
\end{figure*}
\begin{figure}[b!]
\centering
 \includegraphics[width=.4\textwidth]{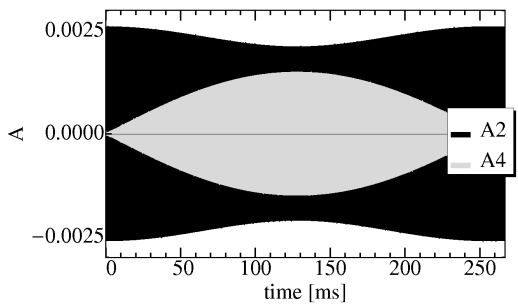}
\caption{The envelopes of the amplitudes of the second and fourth eigenmode
         coefficients $A_i$
         obtained for model 2 and an initial displacement corresponding to
         the second mode with a surface amplitude of $\xi_{2}=10~{\rm m}$.
         Note that the {\em excitation} of mode $i=4$ is so strong that it
         leads to a visible temporary decrease in $A_2(t)$.
        }
\label{fig: A2A4}
\end{figure}

During the evolution, we measure the presence of a given mode $i$
in the form of the associated eigenmode
coefficient $A_i(t)$. In practice, we find these coefficients to
oscillate periodically; cf.~Figures~\ref{fig: Ai_case1} and
\ref{fig: A2A4}. It is convenient, therefore, to use the resulting oscillation
maximum $A_i=\max | A_i(t) |$
as an overall measure of the degree of {\em excitation} of that
mode in the entire evolution. The intuitive interpretation of
the resulting maxima is greatly helped by their one-to-one relation
with the corresponding displacement of the stellar surface.
We recall for this purpose Eq.~(\ref{eq: zeta_expansion}). By setting
$A_k=1$ and $A_i=0$ for all $i\ne k$ in that relation, we can evaluate
the surface displacement $\zeta(x_{\rm S})$ and, thus,
$\xi(x_{\rm S})$ corresponding to eigenmode $k$ with unit amplitude
$A_k$. In Table \ref{tab:amplitude}, we list the resulting
surface displacements for both stellar models studied in this section.
It becomes clear from the table that we expect to deal
with numerically small values of the coefficients $A_k \ll 1$.


In the remainder of this section we will analyze in detail how initial
data given in the form of one single eigenmode gives rise to the
{\em excitation} of other modes.
\subsection{Model 1: A $\gamma=2$ polytrope}

\subsubsection{Exciting the fundamental mode}\label{sec1}

We first focus on the {\em coupling} of eigenmodes in the case of model 1,
a $\gamma=2$ polytrope which is stable against (linear) radial perturbations.
Our first analysis is based on an initial displacement of the fluid
elements given by the profile of the fundamental eigenmode $j=1$.
The simulations last for a physical time of $5~{\rm ms}$
and the
amplitude of the initial data is varied between $10$ cm and $70$ m.
For illustration, we show in Fig.~\ref{fig: Ai_case1} the time evolution
of the eigenmode coefficients $A_1,\ldots,A_4$ obtained for an initial
surface displacement $\xi_1 = 10~{\rm m}$.
Higher-order
modes show a similar behavior with decreasing amplitude. Note
that the coefficients $A_i$ corresponding to weakly {\em excited} modes
such as $i=3$ and $i=4$ in the figure do not oscillate around zero
but reveal an offset. This effect is not surprising bearing in
mind that the sinusoidal dependence of the eigenmode oscillations
is a direct consequence of the separation of variables in
Eq.~(\ref{eigenmode}) according to $\zeta(t,x)=\zeta_i(x) e^{i\omega_i t}$.
In the linear limit we expect an eigenmode to describe a stellar
oscillation around the equilibrium configuration. In the present
case, however, the weak eigenmodes have to be viewed as perturbations
of a time dependent background, namely the equilibrium configuration plus
any strongly {\em excited} eigenmodes. The observed offset thus arises as
one of the nonlinear effects of the system. Indeed, we empirically find
that it increases quadratically with the amplitude of the initial
data and disappears as we reduce the initial displacement of the fluid
elements to zero.

The {\em excitation} of higher-order modes inevitably corresponds
to a flow of energy away from the initially present mode.
Whereas this effect is too small in the example
of Fig.~\ref{fig: Ai_case1} to be noticeable in the amplitude
$A_1(t)$, the strong {\em coupling} between modes 2 and 4
in the case of model 2
in Fig.~\ref{fig: A2A4} demonstrates a strong amplitude modulation
of the mode coefficients. We will discuss this
exceptionally strong {\em coupling} in more detail in the context of
resonance effects in Sec.~\ref{sec:resonance} below.


As mentioned above, we use the maxima for each of the $|A_i(t)|$
to determine the degree of {\em excitation} of the individual modes.
The resulting values for the first six eigenmodes
are shown in
Fig.~\ref{fig:eigen1}. We have ignored higher-order
modes because
they are very weakly {\em excited} and therefore subject to uncertainties
larger than the limit discussed in Sec.~\ref{sec: code_tests}.

The results in the figure show that
the amplitude of the fundamental eigenmode
grows linearly with the initial amplitude $\xi_1$, as expected.
In contrast, all other eigenmode coefficients exhibit
a quadratic dependence on $\xi_1$
\begin{equation}
 A_i = c_{i,1} (\xi_1)^2 ~\text{.}
\end{equation}

The expansion coefficients $c_{i,1}$ obtained from regression
are listed in Table \ref{tab:c1}. Throughout the entire
amplitude range we do not observe any signs of a transition of
the power laws from quadratic to higher order. This indicates
that the nonlinear interaction is dominated by leading-order
effects. This changes when we consider
different initial data.

\subsubsection{Exciting higher modes}\label{sec2}
\begin{figure*}[t!]
\centering
 \includegraphics[width=.45\textwidth]{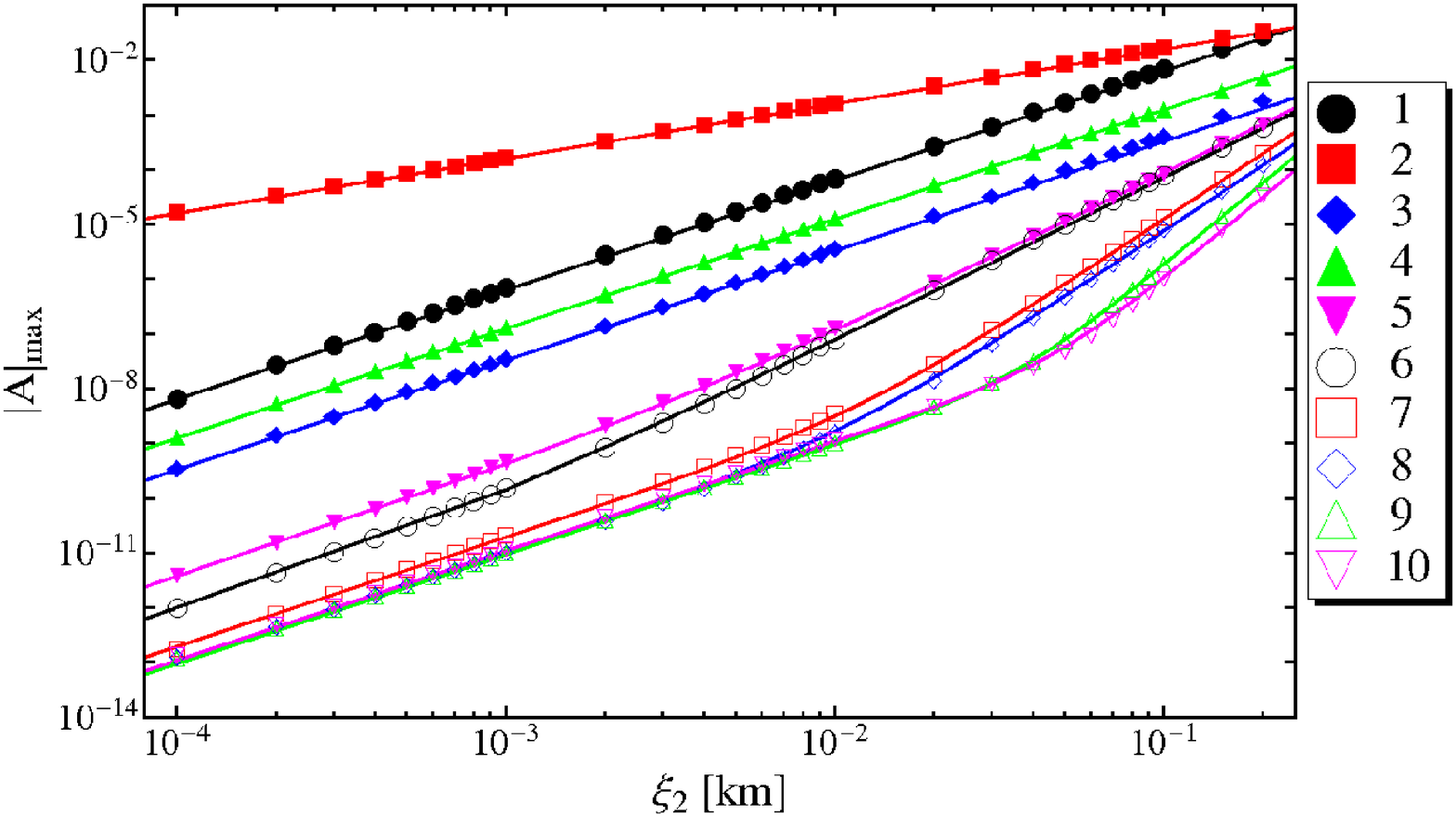}
 \includegraphics[width=.45\textwidth]{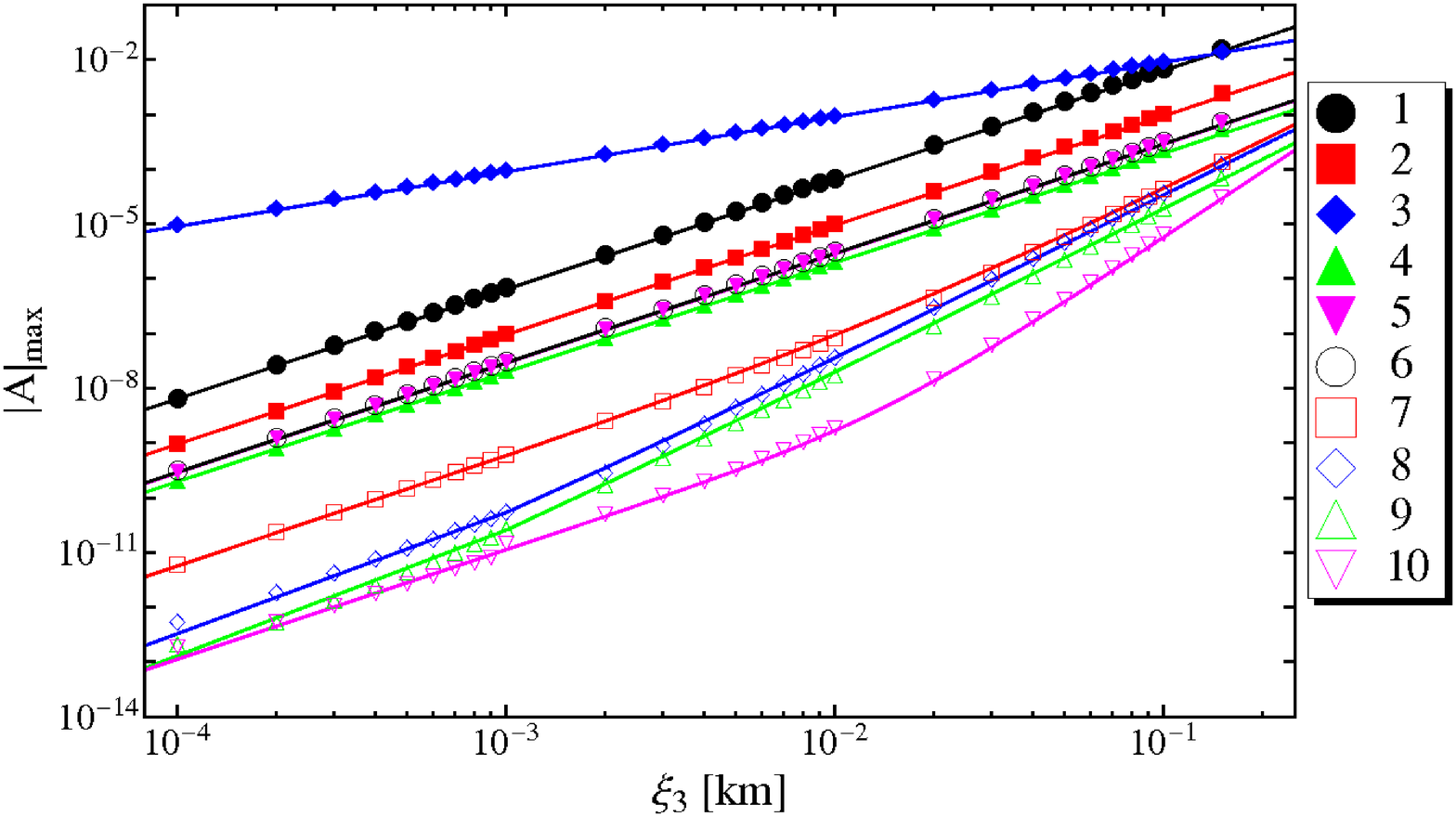}
\caption{The maximal eigenmode coefficients of the first ten eigenmodes
         as functions of the amplitude of the initial perturbation in
         form of the second (left panel) and the third (right panel) mode.
        }
\label{fig:eigen23}
\end{figure*}

\begin{table*}[t]
\begin{center}
\begin{tabular}{c| c | c || c | c | c}
\hline\hline
Mode  &  $~\xi(x_{\rm S})~[{\rm km}]$ model 1 &  $~\xi(x_{\rm S})~[{\rm km}]$ model 2 &Mode  &  $~\xi(x_{\rm S})~[{\rm km}]$ model 1 &  $~\xi(x_{\rm S})~[{\rm km}]$ model 2 \\
\hline
1  &  2.15   &  1.54 & 6  &  28.64  &  14.45\\
2  &  6.31   &  3.84 & 7  &  35.47  &  17.43\\
3  &  11.08  &  6.29 & 8  &  42.74  &  20.51\\
4  &  16.41  &  8.88 & 9  &  50.42  &  23.70\\
5  &  22.27  &  11.60&10  &  58.49  &  26.98\\
\hline\hline
\end{tabular}
\end{center}
\caption{Surface displacement $\xi(x_{\rm s})$ corresponding to eigenmode
         $k$ with unit amplitude $A_k=1$ for the first two stellar models
         of Table \ref{tab:models}.
        }
\label{tab:amplitude}
\end{table*}
\begin{figure}[t!]
\centering
 \includegraphics[width=.48\textwidth]{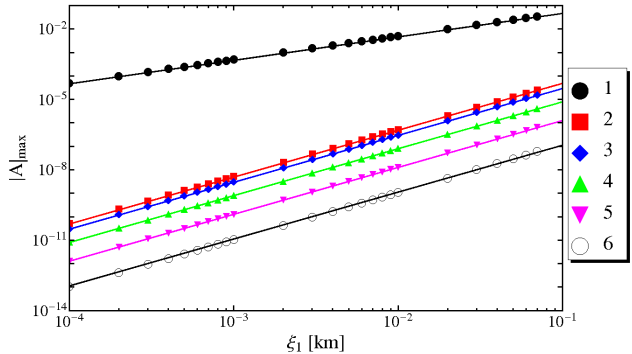}
\caption{The maximal eigenmode coefficients of the first six eigenmodes
         as functions of the amplitude of the initial perturbation in
         form of the fundamental mode. The curves represent a linear
         function in $\xi_1$ for the fundamental mode and quadratic
         power laws for all other modes.
        }
\label{fig:eigen1}
\end{figure}
\begin{table}[t!]
  \centering
  \begin{tabular}{l c c c c c }
    \hline\hline
     $i$      &2&3&4&5&6 \\ \hline
    $c_{i,1}$$[{\rm km}^{-2}]$\hspace{-1mm}&\hspace{-3mm}  $4.9\times 10^{-3}$&\hspace{-1.5mm}$2.9\times 10^{-3}$&\hspace{-1.5mm}$7.9\times 10^{-4}$&\hspace{-1.5mm}$1.3\times 10^{-4}$&\hspace{-1.5mm}$1.1\times 10^{-5}$ \\
  \hline\hline
  \end{tabular}
  \caption{The quadratic expansion coefficients for the {\em mode coupling}
           between the fundamental and other modes.
          }
  \label{tab:c1}
\end{table}
Next, we consider initial perturbations in the form of the second
and the third eigenmode, respectively. The corresponding plots, now
up to and including mode 10, are shown in
Fig.~\ref{fig:eigen23}. As before the eigenmode coefficient of the
initially present mode shows linear behavior. For sufficiently small
amplitudes of the initial displacement, we also observe the quadratic
power law for all modes not present in the initial data.
For larger perturbations, however, these show a clear transition to
higher-order power laws.
This transition demonstrates a significant contribution of {\em couplings}
beyond the leading-order terms and we will refer to this regime as
the {\em moderately nonlinear} regime as opposed to the
{\em weakly nonlinear} regime where quadratic {\em coupling} terms dominate.
The details of the transition such as the threshold amplitude
of the initial data and the relative significance of the higher-order terms 
depend on the individual mode under consideration as well
as on the choice of initial data.
We empirically investigate the transition by fitting the eigenmode
coefficients according to
\begin{equation}
   A_i = c_{i,j} (\xi_j)^2 + d_{i,j} (\xi_j)^3 + e_{i,j} (\xi_j)^4 + f_{i,j} (\xi_j)^5
         + \dots ~\text{.}
   \label{eq: expansion}
\end{equation}
In practice, we find this series to be dominated by a subset of the terms
on the right-hand side and we explicitly set subdominant coefficients
to zero. The resulting coefficients are listed in
Table \ref{tab:c23} and the corresponding fits are shown
as the lines
in Figs.~\ref{fig:eigen1}, \ref{fig:eigen23} and \ref{fig:Model2eigen2}.
While there appears to be a tendency for the expansion coefficients
$c_{i,j}$, $d_{i,j},\ldots$ to decrease for larger mode numbers $i$,
there are exceptions to this rule. Consider, for example,
$c_{3,2}$ and $c_{4,2}$ which demonstrate that the second mode
couples more strongly to mode $4$ than to mode $3$. This is
corroborated by the data for mode $3$ (filled diamonds)
and $4$ (filled upward triangle) in the
left panel of Fig.~\ref{fig:eigen23}. It is interesting to note
in this context that the frequency ratio of modes $2$ and $4$
is close to 1:2. We will return to the issue of resonance phenomena
in Sec.~\ref{sec:resonance}.

Another feature visible both in the expansion coefficients in
Table \ref{tab:c23} and the curves in Fig.~\ref{fig:eigen23} is
that the modes {\em excited} due to nonlinear effects
form groups of similar polynomial behavior. For $j=2$
they form pairs and for $j=3$
triplets. As a function of the initial amplitude of mode 2 (mode 3),
for example, modes 5 and 6 (modes 7, 8 and 9)
show a transition to a cubic power law and modes
7 and 8 (modes 10, 11 and 12)
a transition to a fourth-order power law.
See also the block diagonal structure in
Table \ref{tab:c23} listing the coefficients
$d_{i,2}$, $e_{i,2}$ and $f_{i,2}$ (coefficients
$d_{i,3}$ and $e_{i,3}$).

\begin{table*}
  \centering
  \begin{tabular}{r|cccc|cccc}
    \hline \hline
    $i$  &  $c_{i,2}$  &  $d_{i,2}$  &  $e_{i,2}$  &  $f_{i,2}$  &  $c_{i,3}$  &  $d_{i,3}$  &  $e_{i,3}$  \\
         &[km${}^{-2}$]&[km${}^{-3}$]&[km${}^{-4}$]&[km${}^{-5}$]&[km${}^{-2}$]&[km${}^{-3}$]&[km${}^{-4}$]\\
    \hline
    1    &  0.63       &  0          &  0          &  0          &  $0.64$     &  $0$        &  $0$        \\
    2    &   $-$       &  $-$        &  $-$        &  $-$        &  $0.095$    &  $0$        &  $0$        \\
    3    &  $0.033$    &  $0$        &  $0$        &  $0$        &   $-$       &  $-$        &  $-$        \\
    4    &  $0.12$     &  $0$        &  $0$        &  $0$        &  $0.020$    &  $0$        &  $0$        \\
    5    &  $4.1\times 10^{-4}$ & $0.078$ &  $0$   &  $0$        &  $0.029$    &  $0$        &  $0$        \\
    6    &  $1.3\times 10^{-4}$ & $0.069$ &  $0$   &  $0$        &  $0.029$    &  $0$        &  $0$        \\
    7    &  $1.9\times 10^{-5}$ & $0$&  $0.118$    &  $0$        &  $5.8\times 10^{-4}$ & $0.040$ & $0$    \\
    8    &  $9.4\times 10^{-6}$ & $0$&  $0.073$    &  $0$        &  $4.9\times 10^{-5}$ & $0.034$ & $0$    \\
    9    &  $9.3\times 10^{-6}$ & $0$&  $0$        &  $0.172$    &  $2.3\times 10^{-5}$ & $0.019$ & $0$    \\
   10    &  $1.1\times 10^{-5}$ & $0$&  $0$        &  $0.101$    &  $1.1\times 10^{-5}$ & $0$     & $0.057$\\
   11    &  $-$        &  $-$        &  $-$        &  $-$        &  $1.1\times 10^{-5}$ & $0$     & $0.036$\\
   12    &  $-$        &  $-$        &  $-$        &  $-$        &  $1.2\times 10^{-5}$ & $0$     & $0.020$\\
    \hline \hline
  \end{tabular}
\caption{Expansion coefficients for model 1 obtained from fitting
         Eq.~(\ref{eq: expansion}) to the data points for initial
         data in the form of mode $j=2$ and $j=3$, respectively.
        }
\label{tab:c23}
\end{table*}
To summarize our findings, we observe a weakly
nonlinear regime in which the amplitude of secondary modes
grows quadratically with the amplitude of the initial displacement.
In the case of $j=2$ or $j=3$ and
sufficiently large amplitudes, the mode {\em excitation} exhibits a transition
to higher-order power laws. Secondarily {\em excited} modes form
multiplets; for initial data in the form of mode $j>1$,
the {\em excitation} of modes $i=j+1,...,2j$ depends quadratically on the
initial amplitude, modes $i=2j+1,...,3j$ show a transition to a third-order power 
law, modes $i=3j+1,...,4j$ a transition to a fourth-order
power law and so on. Finally, higher-order modes appear to have a
significantly stronger tendency to transfer energy to lower-order modes
than the other way around. The latter observation was also indicated
by Papadopoulos and Sopuerta's \cite{Papadopoulos2001}
study of black-hole oscillations and suggested to
explain the robustness of black-holes to strong deformations.

\subsection{Model 2: A stiffer equation of state}\label{sec4}
\begin{figure*}[t!]
\centering
 \includegraphics[width=.47\textwidth]{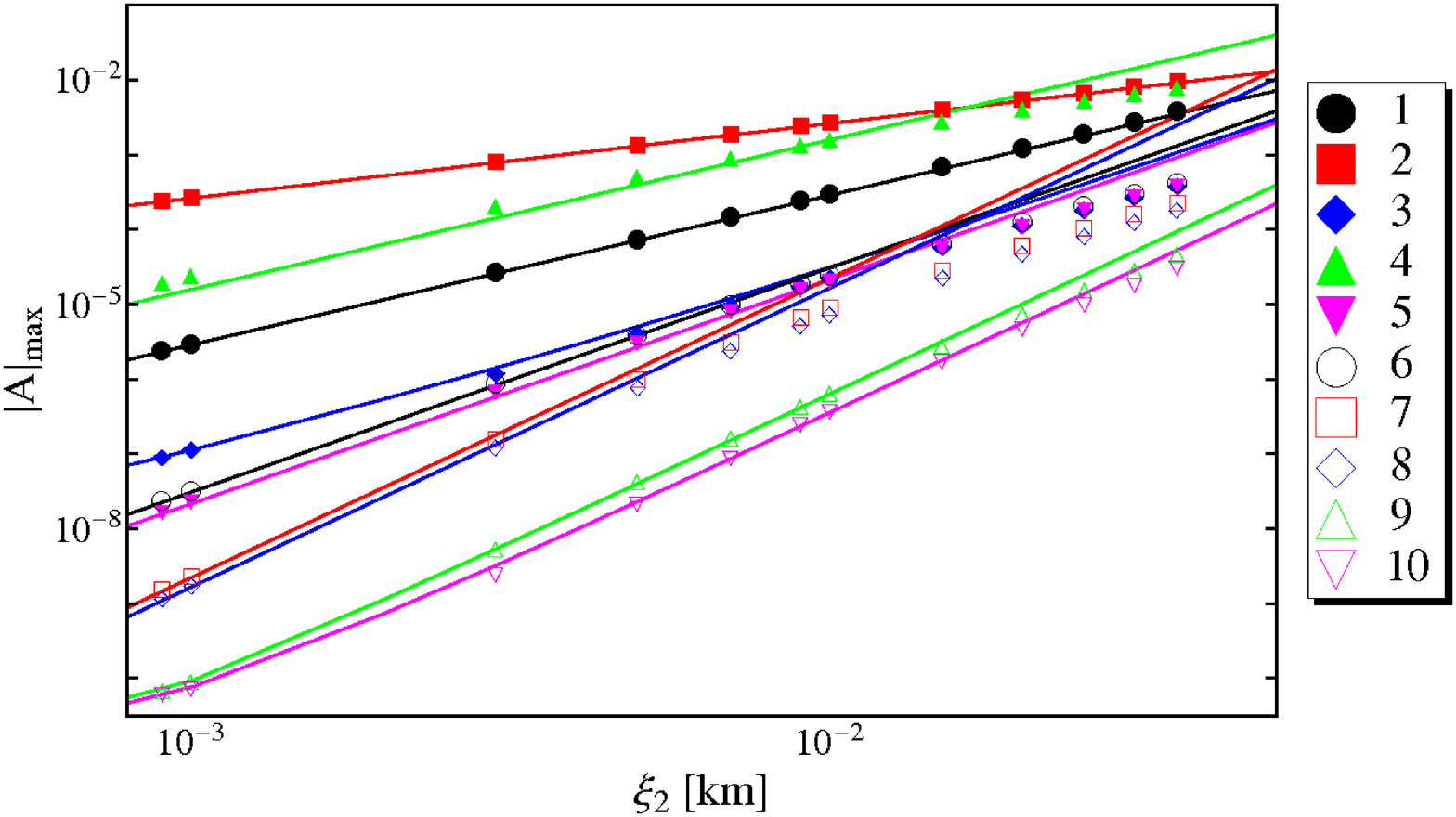}
 \includegraphics[width=.47\textwidth]{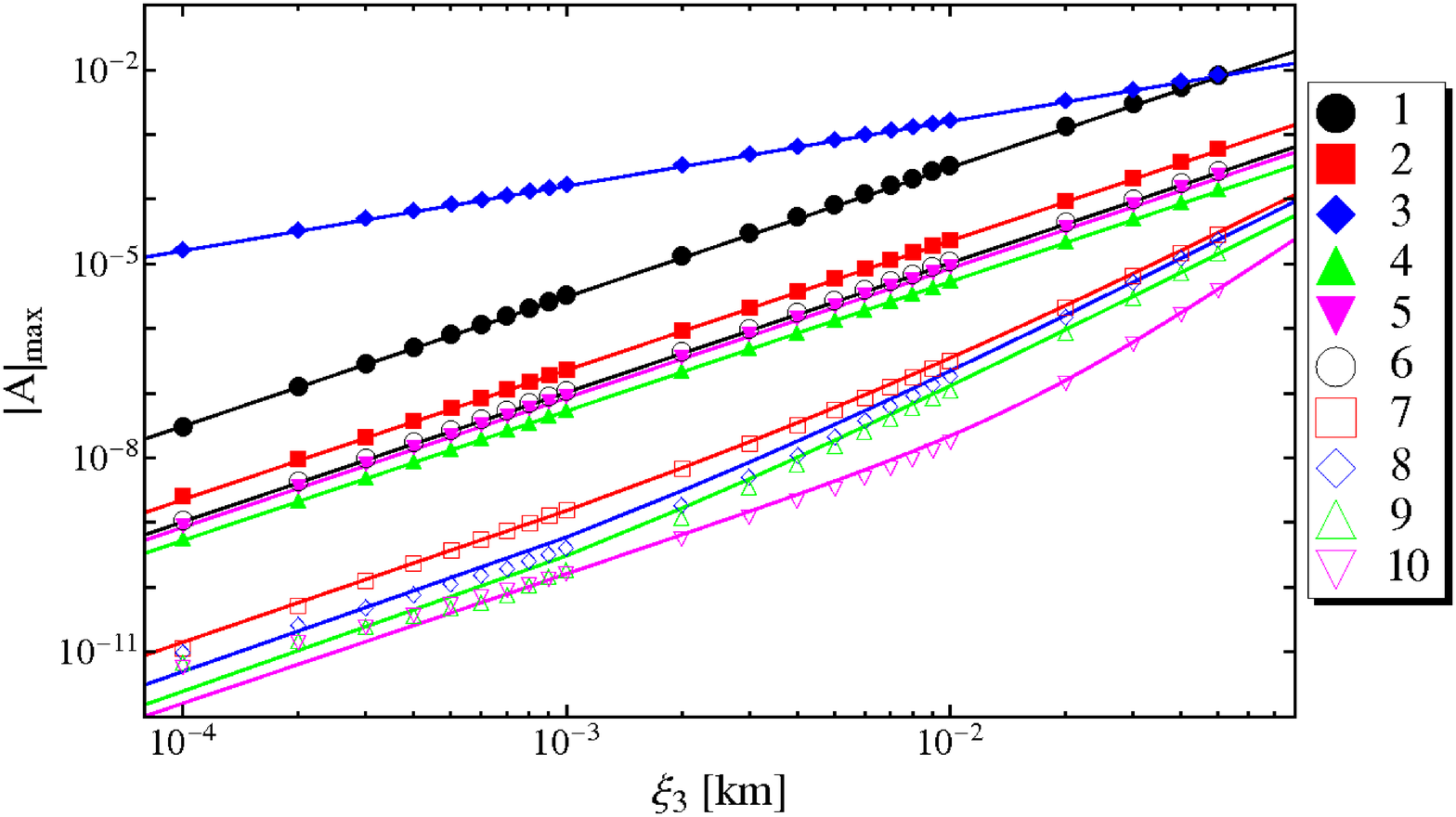}
\caption{The maximal eigenmode coefficients of the first ten eigenmodes
         as functions of the amplitude of the initial perturbation in
         form of the second eigenmode (left panel) or
         the third eigenmode (right panel) as obtained for model 2.
        }
\label{fig:Model2eigen2}
\end{figure*}

We now turn our attention to model 2 of Table~\ref{tab:models}. We want to
assess to what extent the observations in the previous section
depend on the equation of state of the stellar model.
As before, we consider three scenarios where we prescribe initial
data in the form of mode $j=1$, $2$ or $3$ and measure the
degree of {\em excitation} of the other modes up to and including $i=10$ due
to nonlinear effects.
It is reassuring to see that using this model we qualitatively
reproduce most of the general features discussed above. For comparison with the
results for model 1,
we show in Tables \ref{tab:model4_c1} and \ref{tab:model4_c23}
the expansion coefficients
and in Fig.~\ref{fig:Model2eigen2} the eigenmode coefficients
resulting from initial data in the form of eigenmode $j=2$ or $3$.

While confirming the overall behavior observed for model 1, however,
the analysis of model 2 reveals additional complications arising
from two effects which were barely
visible in the analysis of model 1: saturation and resonance.

\begin{table}[b!]
  \centering
  \begin{tabular}{c | *{5}{c} }
    \hline\hline
     i      &2&3&4&5&6\\ \hline
    $c_{i,1}~[{\rm km}^{-2}]$         &$0.032$&$0.011$&$1.5\cdot 10^{-3}$ &$7.6\cdot 10^{-5}$ &$8.9 \cdot 10^{-6}$\\
    \hline\hline
  \end{tabular}
\caption{The quadratic expansion coefficients $c_{i,1}$ for the {\em mode
         coupling} between the fundamental and other modes for the
         stiffer neutron star model.
        }
\label{tab:model4_c1}
\end{table}
\begin{table*}[t!]
  \centering
  \begin{tabular}{r|cccc|cccc}
    \hline \hline
    $i$  &  $c_{i,2}$ [km$^{-2}$] &  $d_{i,2}$ [km$^{-3}$] &  $e_{i,2}$[km$^{-4}$]  &  $f_{i,2}$[km$^{-5}$]  &  $c_{i,3}$ [km$^{-2}$] &  $d_{i,3}$ [km$^{-3}$] &  $e_{i,3}$ [km$^{-4}$] \\
    \hline
    1    &  2.8        &  0          &  0          &  0          &  $3.14$     &  $0$        &  $0$        \\
    2    &   $-$       &  $-$        &  $-$        &  $-$        &  $0.22$     &  $0$        &  $0$        \\
    3    &  $0.09$     &  $0$        &  $0$        &  $0$        &   $-$       &  $-$        &  $-$        \\
    4    &  $24$       &  $0$        &  $0$        &  $0$        &  $0.053$    &  $0$        &  $0$        \\
    5    &  $2.1\times 10^{-3}$ & $21   $ &  $0$   &  $0$        &  $0.084$    &  $0$        &  $0$        \\
    6    &  $5.0\times 10^{-4}$ & $31   $ &  $0$   &  $0$        &  $0.099$    &  $0$        &  $0$        \\
    7    &  $2\times  10^{-6} $ & $0$&  $2210 $    &  $0$        &  $1.3\times 10^{-3}$ & $0.22$  & $0$    \\
    8    &  $<2\times 10^{-6} $ & $0$&  $1650 $    &  $0$        &  $4.1\times 10^{-4}$ & $0.18$  & $0$    \\
    9    &  $<2\times 10^{-6} $ & $0$&  $62.8$     &  $0$        &  $2.0\times 10^{-4}$ & $0.11$  & $0$    \\
   10    &  $<2\times 10^{-6} $ & $0$&  $35.2$     &  $0$        &  $1.5\times 10^{-4}$ & $0$     & $0.52$ \\
    \hline \hline
  \end{tabular}
\caption{Expansion coefficients for model 2 obtained from fitting
         Eq.~(\ref{eq: expansion}) to the data points for initial
         data in the form of mode $j=2$ and $3$, respectively.
        }
\label{tab:model4_c23}
\end{table*}
\subsection{Saturation}\label{sec5}

We first discuss the saturation effect observed in the analysis of model 2.
Consider for this purpose the left panel of Fig.~\ref{fig:Model2eigen2}
which shows the {\em excitation} of modes resulting for $j=2$.
We first notice a relatively strong
{\em excitation} of mode 4; compare the filled upper triangles in this
plot with their counterparts in Fig.~\ref{fig:eigen23} for model 1.
Indeed, the {\em excitation} of mode 4 is so strong that the resulting
power law fit crosses the curve for mode 2 at initial amplitudes
just above $10^{-2}~{\rm km}$. Note, however, how the actual mode
coefficients for mode 4 (filled upper triangles) start deviating from
the polynomial fit at such large amplitudes; the actual {\em excitation} of
mode 4 is weaker than expected from extrapolation of the
polynomial fit and the
coefficients $|A_4|_{\rm max}$ remain below those of the
initially present mode $|A_2|_{\rm max}$. It appears that at
sufficiently large amplitudes, mode 4 starts acting as a significant
source of {\em excitation} itself and transfers energy to other
modes via nonlinear {\em coupling}.
The equilibrium reached in this way sets a limit for the
growth in amplitude of mode 4.
On closer investigation of the left panel of Fig.~\ref{fig:Model2eigen2},
we notice a similar behavior for modes 5 to 8. These modes show saturation
at smaller amplitudes which is compatible with our general
observation that higher-order modes tend to transfer energy to
lower-order modes more efficiently than the other way round.

We conclude from this observation
that for the present example the assumption of
two-mode {\em coupling} would oversimplify the situation. An accurate
modeling of the chain of {\em excitation} whereby energy is
transferred from mode 2 via mode 4 to other modes requires us to
take into account higher-order effects in the {\em mode coupling} as
provided, for example, by our fully nonlinear numerical treatment.

\subsection{Resonance effects}\label{sec:resonance}
Resonance is a well-known phenomenon arising in the context of the
forced oscillator.
Following Van Hoolst \cite{VanHoolst1996}, we can model
the stellar oscillations
as a system of forced oscillators. The
 time evolution of the oscillation amplitudes
is then determined by
\begin{equation}
  \frac{d^2A_i}{dt^2} + \omega_i^2 A_i = \frac{a_0}{2}
         +  \sum_n (a_n \sin n \Omega t + b_n \cos n\Omega t)~\text{.}
  \label{eq: hoolst}
\end{equation}
Here $\Omega$ is the frequency and the $a_n$ and $b_n$ are the amplitudes
of the driving forces. Assuming for simplicity that the $b_n$ vanish,
Eq.~(\ref{eq: hoolst}) has the analytic solution
\begin{equation}
   A_i(t) = \sum_n \frac{a_n}{\omega_i^2 - (n\Omega)^2} \sin n\Omega t
            \hspace*{1cm} n~\epsilon~ \mathcal{N}~\text{.}
\end{equation}
\begin{table}[t!]
  \begin{tabular}{c|c|c||cc|cc}
    \hline \hline
    &Model 1&Model 2&\multicolumn{2}{c}{Model 1}& \multicolumn{2}{c}{Model 2}\\
    Mode $i$  &$\omega_i$ [kHz]&$\omega_i$ [kHz]&$\omega_{2i}/\omega_{i}$& $\omega_{3i}/\omega_{i}$&$\omega_{2i}/\omega_{2}$& $\omega_{3i}/\omega_{i}$ \\

    \hline
    1 & 10.908 & 11.197 & -	& -	& -	& -\\
    2 & 35.299 & 47.526 & 1.975	& 2.890	& 2.000	& 2.939\\
    3 & 53.011 & 72.024 & 1.925	& - 	& 1.940	& -\\
    4 & 69.711 & 95.065 & 1.919 & -	& 1.931 & -\\
    5 & 85.985 & 117.527 & -	& -	& -	& -\\
    6 & 102.026& 139.694 & -	& -	& -	& -\\
    7 & 117.935& 161.685 & -	& -	& -	& -\\
    8 & 133.756& 183.570 & -	& -	& -	& -\\
  \hline\hline
  \end{tabular} \\[5pt]
  \caption{The frequencies and the most promising resonance factors
           $\omega_{2i} / \omega_{i}$ or $\omega_{3i} / \omega_{i}$ for some of the
           eigenmodes of model 1 and model 2.
          }
  \label{tab:freq}
\end{table}
In our case, the driving terms are typically dominated by the single
eigenmode $j$ present in the initial data
and therefore $\Omega=\omega_j$. This mode should excite with
particular efficiency those modes which have an eigen frequency
close to an integer multiple of $\omega_j$. In Table \ref{tab:freq}
we show the frequencies and corresponding ratios for some modes
of model 2 whose frequencies have ratios close to integer values.
Resonance occurs most conspicuously between modes 2 and 4 and we
have already noted the strong {\em excitation} of mode 4 in the
left panel of Fig.~\ref{fig:Model2eigen2}. The effect is not as
dramatic in the case of modes 6 and 8, but we still observe a
preferred {\em excitation} of these modes compared with model 1;
compare, for example, their {\em excitation} with that of
mode 3 and the corresponding
results in the left panel of Fig.~\ref{fig:eigen23}.
We also note a small deviation in the transition to different
power laws in the moderately nonlinear regime. Consider
modes $i=9,\,10$ in the case of $j=2$ in
Table \ref{tab:model4_c23}. According to the rule, the
{\em excitation} of these modes should exhibit a transition to
fifth order in $\xi_2$, but instead we observe a fourth-order dependence. 
We believe this to be a consequence of
the strong {\em excitation} of mode $4$, so that our approximation
of a single strong mode driving the {\em coupling} is no longer
valid.

In all cases of {\em mode coupling}, we observe a periodic transfer
\begin{figure}[b!]
\centering
 \includegraphics[width=.47\textwidth]{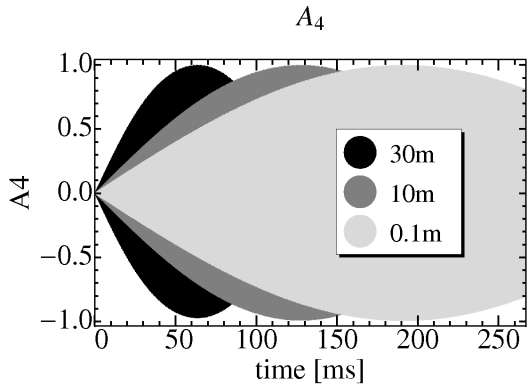}
\caption{The evolution of the fourth eigenmode coefficient for
         different perturbation amplitudes.
        }
\label{fig:modulation}
\end{figure}
of energy back and forth between the eigenmodes involved. This manifests
itself in a modulation of the oscillation amplitude as apparent, for
example, in Fig.~\ref{fig: Ai_case1}, in particular the upper right panel.
Such amplitude modulation is also present in the case of strong resonance.
In contrast to ``normal'' {\em coupling} of eigenmodes, however,
we further observe in cases of strong resonance a modulation period
which depends sensitively on the amplitude of the initial data.
\begin{table}[t!]
  \centering
  \begin{tabular}{c| c}
    \hline\hline
    initial amplitude  & frequency $\Omega$  \\
    $[m]$ & $[{\rm Hz}]$\\
    \hline
    0.1&$8.21$\\
    1&$8.27$\\
    10&$12.29$\\
    20&$18.98$\\
    30&$25.30$\\
    \hline\hline
  \end{tabular}
  \caption{This table presents the frequencies obtained by
           fitting the envelopes of the eigenmode coefficients
           evolution with sinusoidal functions.
          }
\label{tab:sinus}
\end{table}
We illustrate this in Fig.~\ref{fig:modulation} which displays
the coefficient $A_4(t)$ obtained for model 2 and initial data
in the form of the second eigenmode with different amplitudes.
As is evident in the figure, larger amplitudes result in shorter
modulation periods. The resulting numerical values of the modulation
frequency are shown in Table \ref{tab:sinus} and are well approximated
by the linear relation
\begin{equation}
  \Omega_{\rm mod}(\xi_2) = 7.55 ~{\rm Hz} + 577 ~{\rm Hz} \frac{\xi_2}{\rm km}~\text{.}
  \label{eq:Omfit}
\end{equation}
In the perturbative limit $\xi_2\rightarrow 0$ we
obtain a modulation frequency of
$\Omega_{\rm mod}=7.55~{\rm Hz}$. A detailed interpretation in
the context of coupled harmonic oscillators \cite{VanHoolst1996}
is beyond the scope of this
paper, largely because it is highly nontrivial to calculate the
coupling constants. We note, however, the close resemblance between
the measured modulation frequency and the difference between the
involved eigenmodes' frequencies
\begin{equation}
  \Omega_{\rm mod}\approx \frac{\delta\omega}{2} =
       \frac{\omega_4 - 2\cdot \omega_2}{2}\approx 6.5~{\rm Hz}~\text{.}
  \label{eq:Ommod}
\end{equation}

Note that $\delta \omega$ is orders of magnitude below
the individual frequencies $\omega_2$ and $\omega_4$ and therefore
subject to a larger relative error. Within this error, the measured
value agrees with the prediction of Eq.~(\ref{eq:Omfit}).

Leaving a detailed investigation for future work,
we tentatively interpret our observations as follows. The modulation
frequency is closely related to the difference in the frequencies of
the {\em coupling} modes and decreases significantly as we approach
resonance. As has already been discussed in Sec.~\ref{sec5},
nonlinear effects set a limit on the resonance. They further
result in a deviation of the modulation frequency from the
limit of perfect resonance $\Omega_{\rm mod}
\rightarrow 0~{\rm Hz}$.

In comparison, the resonance between modes 2 and 4 of model 1
is less pronounced and we find the modulation frequency to be
independent of the amplitude of the initial data. Indeed,
Eq.~(\ref{eq:Ommod}) predicts  $\delta \omega= 141~{\rm Hz}$
for model 1. This value is an order of magnitude larger than
that obtained for model 2 and appears to be much more robust to
effects of a finite initial amplitude $\xi_2$ as given by the
second term on the right-hand side of Eq.~(\ref{eq:Omfit}).

\section{Further nonlinear effects}
\label{sec: further_effects}

In this section we consider two nonlinear effects which are not directly related
to the {\em coupling} of eigenmodes.
First, we study the stability properties of a stellar model close
to the maximum of the mass-radius relation but located on the unstable
branch. From linear theory, we would expect this model to be unstable
to small perturbations away from its equilibrium configuration.
Second, we investigate the consequences of the vanishing of the speed of
sound at the stellar surface with regard to the formation of discontinuities
near the surface.

\subsection{Stabilization of linearly unstable stars}
\begin{figure}[b!]
\centering
\includegraphics[width=.35\textwidth]{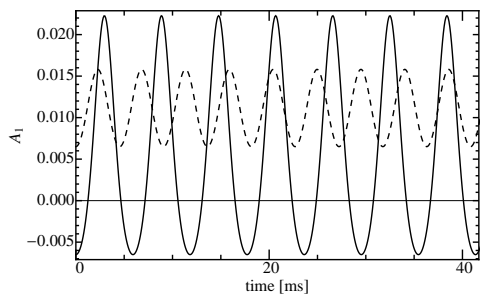}
\caption{The coefficient of the fundamental eigenmode. The dashed
         line corresponds to an initial expansion of the star,
         the solid line to a compression.}
\label{fig:unstable}
\end{figure}
A particularly interesting scenario for the investigation of nonlinear effects
is that of marginally stable neutron stars. It is well known that
neutron stars of sufficient compactness
become unstable with respect to their fundamental
radial oscillation mode \cite{Chandrasekhar1964}. The onset of this
instability occurs at zero frequency of the fundamental mode. At this
point linear perturbation terms cancel in the equations and the
higher-order perturbations become more important.

We study this effect quantitatively for
model 3 of Table \ref{tab:models}. We have already mentioned that this model
is unstable, i.e.~its fundamental oscillation mode has an imaginary
frequency $\omega_1^2 = -779$ Hz$^2$.
We test the prediction of the linearized theory by prescribing initial
data in the form of this fundamental eigenmode using an amplitude of
$10~{\rm m}$. The surprising result of the fully nonlinear time evolution
is shown in Fig.~\ref{fig:unstable}; the star oscillates periodically
over many milliseconds without any sign of instability. We find this
general behavior independent of whether the initial displacement
represents an expansion (dashed curve) or a compression (solid curve)
of the star, even though these two scenarios differ in the details
of the ensuing oscillations. This result demonstrates that
nonlinear effects have the capacity to stabilize a linearly
unstable neutron star. It will be interesting in future work
to study this effect in more detail from an analytic point of view.

In addition to this qualitative difference between perturbative predictions
and the nonlinear evolution, we observe a variety of quantitative
deviations.
Contrary to expectations from linearized theory,
the sign of the initial perturbation of
the star has an impact on the resulting oscillation pattern.
In particular, the frequency and amplitude of the oscillation differ
significantly for the two cases as is apparent in Fig.~\ref{fig:unstable}.  
We further observe an offset
of the eigenmode coefficient indicating that the star is no longer
oscillating symmetrically around its equilibrium position. We have already
identified such an offset as a nonlinear effect in Sec.~\ref{sec1}.
In the present example, however, the offset
is much larger. We further note that the solid curve does not have
sinusoidal shape but is visibly distorted.
\begin{figure}[t!]
\centering
 \includegraphics[width=.35\textwidth]{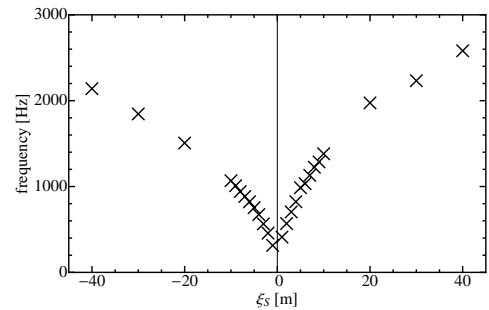}
\caption{The frequency of the fundamental mode as function of the
         initial perturbation. The frequencies have been obtained
         by a Fourier analysis of the nonlinear evolution.
        }
\label{fig:inst}
\end{figure}
Finally, we find that the oscillation frequency of the nonlinear
evolution depends not only on the sign but also on the initial
amplitude of the perturbation. Indeed, the measured values differ
substantially from those predicted by linear theory.
This is illustrated in Fig.~\ref{fig:inst}, where we show the
oscillation frequency as a function of the amplitude in the
range $-40$ m to $+40$ m. As expected intuitively, we obtain larger real
frequencies, i.~e.~larger deviations from the linear prediction,
for larger amplitudes of the initial data. Conversely, we recover
the linear limit and
observe collapse of the stellar model when choosing initial
perturbations of sufficiently small amplitude, in the decimeter range
for the present example.

\subsection{Shock formation at the surface}\label{sec:shock}
\begin{figure*}[t!]
\centering
 \includegraphics[width=.43\textwidth]{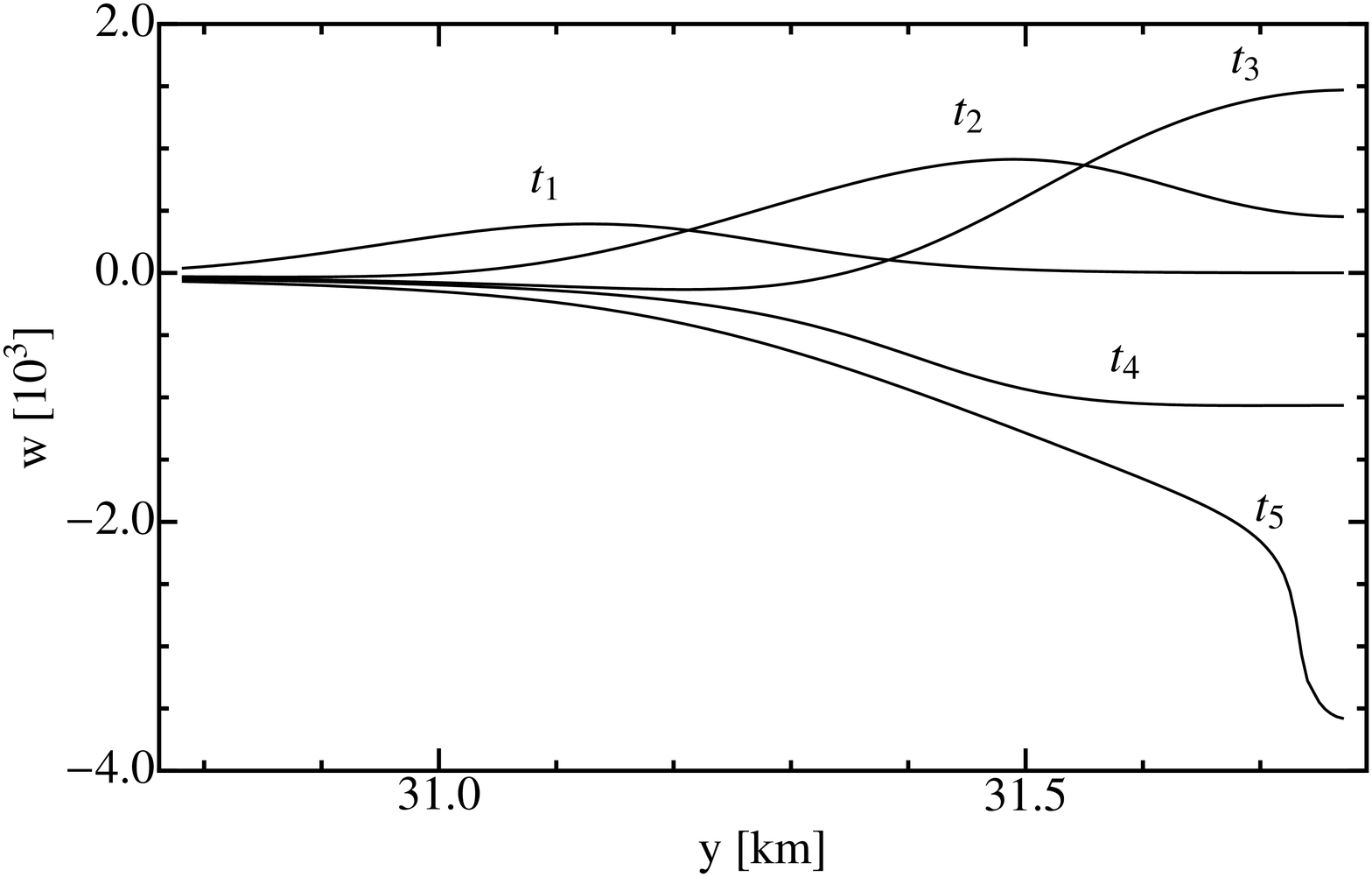}
\includegraphics[width=.43\textwidth]{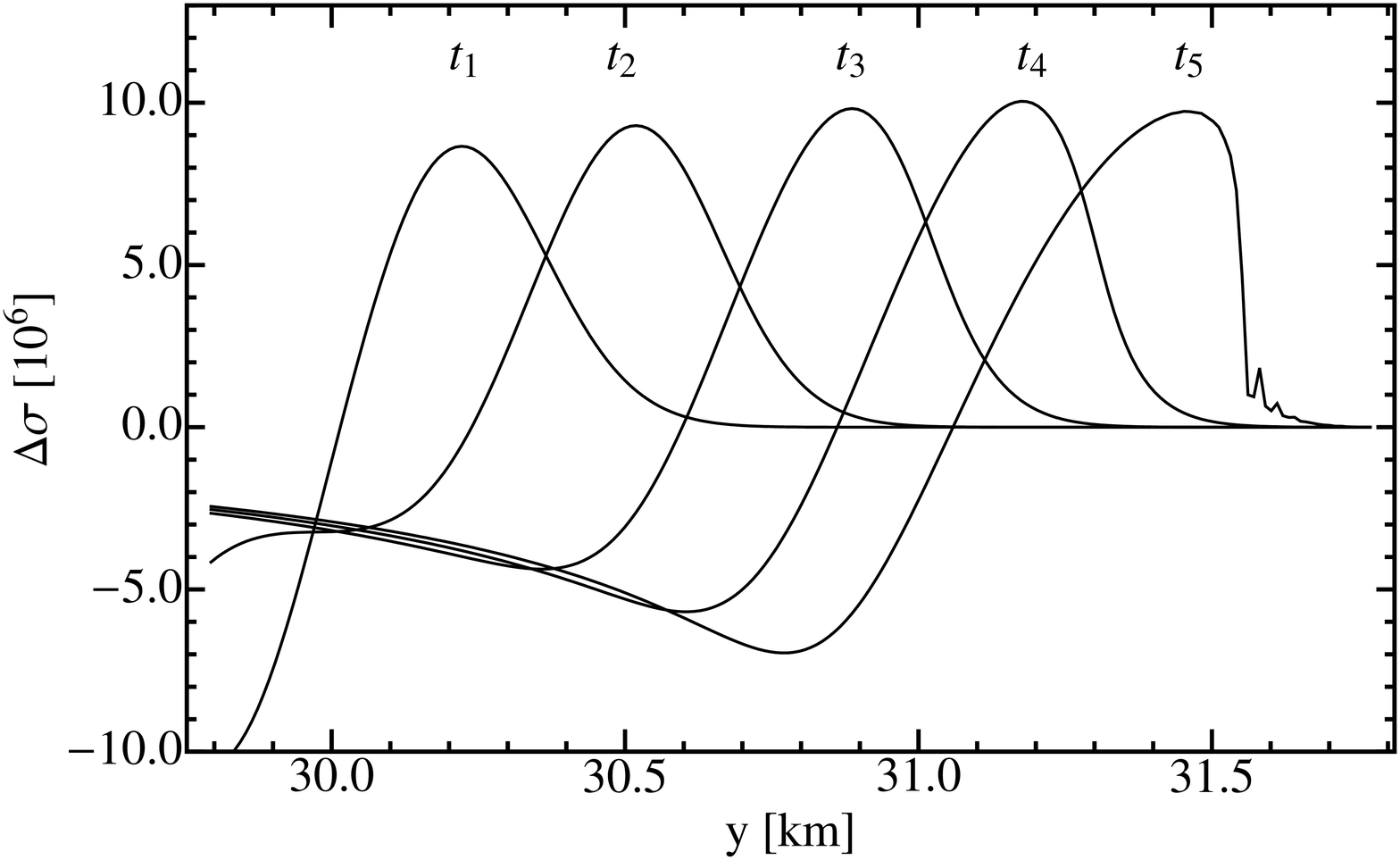}
\caption{Snapshots of the evolution of the velocity $w$ for a moderate
         amplitude $B=2\times10^{-4}$ of the initial perturbation (left panel) and
         of $\Delta \sigma$ obtained for an amplitude $B=10^{-3}$.
        }
\label{fig:pulse}
\end{figure*}

Analysis of the asymptotic structure of the TOV
model combined with a polytropic equation of state (cf.~Sec.~5.2.5
of \cite{Sperhake2001}) shows that the
speed of sound vanishes as $(\bar{r}_{\rm S}-\bar{r})^{1/2}$ at the surface
regardless of the parameters of the polytrope. Here $\bar r_{\rm S}$ is the
radius of the star.
Because of this decrease of the speed of sound, a signal of finite width
propagated towards the surface will be compressed; its tail moves faster
than its head. Naturally one may ask under which circumstances this
gives rise to shock formation.
In order to investigate this question, we consider model 1 of Table
\ref{tab:models}.

First, however, we need to take care of a numerical difficulty
arising in this context. In order to guarantee adequate numerical
resolution of the expected features close to the surface, we switch from the
radial coordinate $x$ to a rescaled radius $y$ related by
\begin{equation}
 dy = \frac{1}{\bar{C}} d\bar{r} ~~\rightarrow ~~ \bar{r}_{,y}
    = \bar{C} ~\text{,}
\end{equation}
where $\bar{C}^2=\frac{\partial \bar{P}}{\partial \bar{\rho}}$
is the speed of sound.
A straightforward calculation shows that the speed of sound measured
in terms of $y$ now approaches a finite value at the surface,
thus eliminating the danger of under resolving
features \footnote{We note that this potential lack of numerical
resolution is not a problem in the evolution of the eigenmodes
discussed above because of their oscillatory character.}.
Even more remarkably, this {\em decompactification}
is achieved by mapping a finite interval in $x$ to a finite
interval in the new variable $y$.
For more details on this procedure we refer the reader to
\cite{Ruoff2000} and \cite{Sperhake2001}.
For the numerical analysis we perturb model 1 by setting the initial
velocity
\begin{equation}
 w(t=0) = B e^{-(y-y_0)^2/b},
\end{equation}
where width and center of the Gaussian pulse are set to $b=0.05~{\rm km^2}$
and $y_0=30~{\rm km}$, respectively. The general behavior of this
pulse during a fully nonlinear time evolution now depends exclusively
on its initial amplitude $B$. In practice, we observe three
different types of behavior.
For sufficiently small amplitudes, the pulse is
reflected at the surface and we do not observe shock formation.
At moderate amplitudes the pulse gets reflected
at the boundary, but a discontinuity in $w$ as well
as $\Delta \sigma$ develops shortly
after reflection. This is shown in the left panel of
Fig.~\ref{fig:pulse} for an amplitude $B=2\times 10^{-4}$.
As we further increase the amplitude, discontinuities develop
in the variables $w$, $\Delta\lambda$ and $\Delta\sigma$
before the pulse reaches the stellar surface.
This scenario is illustrated for the variable $\Delta \sigma$
and using an initial amplitude $B=10^{-3}$
in the right panel of
Fig.~\ref{fig:pulse}.

Shock formation at the stellar surface has recently been studied
in the context of a plane parallel model in Newtonian gravity by
Gundlach and Please \cite{Gundlach2009}. In particular, their
Eq.~(27) provides the minimum amplitude required for shock formation
in terms of amplitude $v_0$ as well as width $\tilde{\sigma}_1$
and initial location $\tilde{\sigma}_0$ of the initial pulse. (We
use a tilde to distinguish their $\tilde{\sigma}$ from our
evolution variable defined in Eq.~(\ref{eq: def_sigma}).) This equation
neglects a factor $1/2$, which we reintroduce for the comparison
with our numerical simulations. The starting point for the following
discussion is therefore
\begin{equation}
 \frac{v_0}{\tilde{\sigma}_1} \lesssim \frac{1}{2}\left(
          \frac{\tilde{\sigma}_1}{\tilde{\sigma}_0}
          \right)^{n+\frac{1}{2}}, \label{eq:shockCarsten}
\end{equation}
where the polytropic index is defined by $\gamma = 1+1/n$.
Translating this expression into the coordinates used throughout
this work, we arrive at the following condition\footnote{This expression
corrects Eq.~(28) in \cite{Gundlach2009} by adding a missing factor
$\sqrt{x_\star/x_0}$.}
\begin{equation}
 D\equiv \frac{v_0 x_\star^{n+1}}{\bar{r}_1^{n+\frac{3}{2}}} \lesssim
     \sqrt{g} =const. ~\text{.} \label{eq:shock}
\end{equation}
Here $x_\star=|r_{\rm S} - \bar{r}_0|$, $\bar{r}_0$ is the position and
$\bar{r}_1$ the characteristic length scale
of the initial perturbation and
$r_{\rm S}$ and $g$ are radius and gravitational acceleration
at the surface of the background model, respectively.
The derivation of relation (\ref{eq:shockCarsten})
does not take into account the precise shape of the wave packet, however,
which may enter as a dimensionless constant.

In order to test their prediction numerically we perform a series
of simulations with Gaussian initial data. As a measure of the
characteristic length scale $\bar{r}_1$ in Eq.~(\ref{eq:shock}),
we choose the {\em full width at half maximum} (FWHM) of the Gaussian.
For our model 1 we have a polytropic index $n=1$
and $\sqrt{g}= \sqrt {M/r_{\rm S}^2}= 0.1337$.
The resulting two-parameter study is summarized in Table \ref{tab:shocks}.
For each combination of width $\bar{r}_1$
and initial position $\bar{r}_0$
we start with a very small amplitude of the initial perturbation
and repeat the simulation at increasingly larger amplitude
until we observe formation of a discontinuity. The resulting
threshold amplitude $v_0$ is then used to evaluate the
left-hand side of Eq.~(\ref{eq:shock}).

\begin{table}[t!]
\centering
\begin{tabular}{c| c c c c}
\hline\hline
$\bar{r}_0$&\multicolumn{4}{c}{$\frac{v_0 |r_{\rm S} - \bar{r}_0|^{n+1}}
{\bar{r}_1^{n+{3/2}}}$ } \\
in [km]&\multicolumn{4}{c}{for initial width $\sqrt{b/2}$ } \\
  &0.2&0.3&0.4&0.5\\
\hline
6.79&0.180&0.196&0.204&0.207\\
7.13&0.168&0.182&0.189&0.191\\
7.45&0.157&0.170&0.176&0.178\\
7.76&0.148&0.160&0.165&0.167\\
8.05&0.141&0.152&0.157&0.158\\
8.32&0.135&0.145&0.150&0.151\\
8.58&0.130&0.140&0.144&0.145\\
8.82&0.126&0.135&0.139&0.140\\
9.05&0.123&0.132&0.135&0.136\\
9.27&0.120&0.129&0.132&0.133\\
9.47&0.118&0.126&0.130&0.130\\
9.66&0.117&0.125&0.128&0.129\\
9.83&0.115&0.123&0.127&0.128\\
9.99&0.114&0.123&0.126&0.127\\
10.13&0.114&0.122&0.126&0.128\\
10.27&0.114&0.122&0.126&0.128\\
10.38&0.114&0.123&0.127&0.130\\
10.46&0.114&0.123&0.129&0.131\\
10.58&(0.115)&(0.125)&(0.131)&(0.134)\\
10.65&(0.116)&(0.127)&(0.134)&(0.139)\\
10.72&(0.118)&(0.130)&(0.138)&(0.141)\\
\hline\hline
\end{tabular}
\caption{The left-hand side of Eq.~(\ref{eq:shock}) for critical
         perturbations that are sufficiently large to generate shocks close
         to the surface.
        }
\label{tab:shocks}
\end{table}

The numerical uncertainties
in this study are dominated by the difficulties in determining
when a discontinuity has actually formed. First,
we have to discretize the amplitude $v_0$ in order to keep
the number of simulations at an acceptable level.
Second,
numerical dissipation may suppress shock formation. Finally,
there is some freedom in choosing a critical gradient to define
a discontinuity. In our simulations we choose the critical value
to be $w_{,y}= 0.05$ km$^{-1}$. Bearing in mind all these difficulties,
we estimate the numerical accuracy of the critical amplitude $v_0$
to be about $10~\%$.
Within these uncertainties, we observe good agreement with the
predictions of \cite{Gundlach2009}. The
obtained values for $D$ are listed in Table \ref{tab:shocks}.

As long as the initial position of the pulse is not too
close to the center
the left-hand side of
Eq.~(\ref{eq:shock}) shows little variation
with the pulse width and initial location.
Bearing in mind that (\ref{eq:shock}) has been derived for
plane parallel geometry and in the Newtonian
approximation, the observed increase in variability of the coefficient
in Table \ref{tab:shocks} at small $x_0$ is not surprising. For radii
close to the surface the determination of the width and
the position of the pulse is limited because the pulse steepens
and deviates significantly from a purely Gaussian profile.
Values obtained in this regime are thus given in parentheses.

In summary, Eq.~(\ref{eq:shock}) as well as the numerical study
illustrate nicely that the significance of nonlinear effects
does not only depend on the amplitude of the evolving feature.
Instead, the length scale on which profiles change is equally
important and may generate significant deviations from
perturbative predictions at comparatively small amplitudes of the
signal.

For completeness we note that the stellar model used in this study
oversimplifies the structure of the stellar surface. A crust expected
to form at the surface of a neutron star is likely to result in
more complex phenomena and requires a more detailed treatment.
As has been mentioned above, however, the primary purpose of our study
is to probe the taxonomy of nonlinear effects in the weakly and
moderately nonlinear regime. We postpone the sophistication
of a realistic neutron star model to future work.
In this spirit, our analysis leads to two main conclusions.
First, the theoretical modeling of neutron star
surfaces requires particular care and, second, the
surface exhibits a rich phenomenology, not always concurrent with
immediately intuitive expectations.

\section{Summary}\label{sec:summary}
In this paper we have investigated a variety of nonlinear
effects associated with radial oscillations of polytropic neutron star models.
We have performed our study in the framework of fully nonlinear
general relativity and have achieved a high level of numerical
precision by formulating the time evolution in terms of
finite deviations from an equilibrium configuration which
in our case is given by a TOV model.

The first nonlinear effect studied in detail is the {\em coupling}
of eigenmodes due to nonlinear effects.
In the absence of a unique decomposition of a fully nonlinear dynamic
model into background and deviations, we
have performed one-parameter studies using as initial data
the equilibrium configuration perturbed in the form
of one single eigenmode profile with varying amplitude.
Our results and their interpretation are to be understood
in the context in this particular
construction of a reference background.
By virtue of the completeness of the eigenmode spectrum,
we are able to measure the {\em excitation} of modes not contained
in the initial data by calculating eigenmode coefficients
from overlap integrals. This analysis has revealed two
qualitatively different regimes depending on the amplitude
of the initial data.
For sufficiently low values, we find the {\em excitation}
of modes to increase quadratically with the amplitude.
In the context of analytic studies based on coupled oscillators,
as for example employed by Van Hoolst \cite{VanHoolst1996},
we interpret this quadratic dependence as a leading-order
{\em coupling} between different modes. In the case of an initial
perturbation given by the fundamental eigenmode, this quadratic
dependence persists over the entire range of initial surface
displacements of up to $70~{\rm m}$. Initial data of higher modes
with sufficient amplitude, however,
result in a transition in the {\em excitation} of modes to higher-order 
power laws. Specifically, our results indicate that secondarily
{\em excited} modes appear in multiplets; initial data of eigenmode $j$
{\em excites} modes $i=j+1,...,2j$ according to a quadratic power law,
modes $i=2j+1,...,3j$ with cubic dependence and so on.
Strong resonance effects may cause some complications, however,
which manifest themselves in deviations from this rule.
This behavior
demonstrates the significance of higher-order
{\em coupling} of eigenmodes. The onset of this moderately nonlinear
regime occurs at amplitudes of the order of a few meters,
but the details appear to vary with the stellar model.
As a general rule, we find higher-order modes to pass energy
to lower-order modes more efficiently than the other way around.

Our results confirm the intuitive expectation that
eigenmodes with an integer frequency ratio interact
particularly efficiently. This resonance manifests itself most
conspicuously in the second and fourth eigenmode of the stiffer
model considered in this study where the frequency ratio is
equal to two within three digits. Initial data in the form of
the second mode can excite the fourth mode with such efficiency
that both amplitudes become comparable. When this happens, we observe
a further complication in the nonlinear behavior: the
strong {\em excitation} of mode 4 results in a significant transfer of
energy to other eigenmodes and its amplitude saturates, i.e.~it stops
growing in accordance with the expected power law. A similar
behavior is observed for other pairs of modes, as, for example,
modes 2 and 6. Overall, resonance appears to be weaker for
larger deviations of the involved frequency ratios from
an integer value.

A particularly fertile ground for the analysis of nonlinear effects
is given by stellar models close to the stability limit of the
mass-radius relation. Instability manifests itself in the vanishing
of the oscillation frequency of the fundamental eigenmode
\cite{Chandrasekhar1964} in which case the
linear (in the deviations) terms in the evolution equations cancel.
The resulting dominance of higher-order terms leads to a variety
of effects. Most notably, our results show that nonlinear effects
have the capacity to stabilize stars expected to be unstable in
linear analysis; initial perturbations lead to a periodic pulsation
of the star instead of the expected gravitational collapse.
Nonlinearity further manifests itself in a visible distortion of
the time dependence of the oscillations away from sinusoidal
character. Also, the oscillation pattern depends on the initial
phase of the perturbation.
It will be interesting to investigate in future work to what
extent this stabilization is a generic feature and how it
depends on the physical properties of the stellar model.

Finally, we have studied the formation of discontinuities near the
surface of the star by evolving Gaussian pulses propagating from the
stellar interior towards the surface.
In particular, we have compared
our numerical findings with analytic predictions by
Gundlach and Please \cite{Gundlach2009} who give threshold
amplitudes for the initial amplitude of the pulse depending
on its initial location in the sense that amplitudes
above this threshold lead to shock formation and those below
do not.
Within the accuracy of the analysis, approximately $10~\%$,
our numerical results confirm the predictions of Gundlach and Please
in the range of validity of their model.

In summary, the precision of the numerical method proposed here
has enabled us to identify a number of nonlinear
effects with no approximation other than that of a rather
simple stellar model. Our results suggest a variety of
extensions for future work. These include the refinement
of the stellar model by including, for example, realistic
equations of state, magnetic fields and, most importantly,
relaxing the condition of spherical symmetry. The main
obstacle for the latter appears to us to be the extension
of the comoving, Lagrangian formulation to higher dimensional problems.
We find a careful treatment of the stellar surface
to be crucial in our study. A possible avenue towards achieving
this goal is the use of {\em level set methods} and {\em
fast marching methods} \cite{Sethian1999} specifically designed
to follow the motion of interfaces in numerical simulations.

\begin{acknowledgments}
We thank Philippos Papadopoulos for various fruitful discussions.

This work was supported by DFG grant SFB/Transregio~7 ``Gravitational
Wave Astronomy'', by grants from the Sherman Fairchild
Foundation to Caltech and by NSF Grant Nos. PHY-0601459 and PHY-0652995.
US acknowledges support from NSF TeraGrid Grant No. PHY-090003.
NA acknowledges support from STFC in the United Kingdom via Grant No.
PP/E001025/1.

\end{acknowledgments}

\appendix

\section{The eigenmode equation}\label{AP:eigenmodeequation}

In the limit of infinitesimally small deviations $\Delta f$ from
the equilibrium variables $\bar{f}$, the time evolution of the star is
governed by the linearized version of the system
(\ref{nonlinear-sigma}) - (\ref{nonlinear-w}).
The combination of these equations into the single, second-order
partial differential Eq. (\ref{eigenmode})
for the rescaled displacement $\zeta$
is discussed in detail in \cite{Misner1973}. The coefficients
 $W$, $\Pi$ and $Q$ in that equation are given by
\begin{eqnarray}
  \Pi &=& \bar{C}^2 (\bar{\rho} + \bar{P}) \frac{\bar{\lambda}^3 \bar{\mu}}
          {\bar{r}^2}  \label{Pi}\\
  W &=& (\bar{\rho} + \bar{P}) \frac{\bar{\lambda} \bar{\mu}^3}{\bar{r}^2}
        \label{eq: W} \\
  Q &=& \frac{\Pi}{\bar{C}^2} \left[ \left(\frac{\bar{\lambda}_{,x}}
        {\bar{\lambda}}\right)^2+\frac{4}{\bar{r}}
        \frac{\bar{\lambda}_{,x}}{\bar{\lambda}}
        - 8\pi \bar{\mu}^2 \bar{P}  \right], \label{eq: Q}
\end{eqnarray}
where the metric function $\bar{\mu}=(1-2\bar{N}\bar{r})^{-1/2}$.

In order to numerically solve the eigenmode equation, we write it as a
first-order system in the variable $\xi$ and make the ansatz
$\xi(t,x)=\xi(x) e^{i\omega t}$.
\begin{eqnarray}
   0&=& \bar{C}^2 \xi_{,x} - B \label{defB}\\
   B_{,x}&=&  - \frac{\bar{C}^2 \bar{r}_{,x}}{\Pi} \left\lbrace \frac{\bar{\lambda}}{\bar{r}^2}
        \left[ \frac{\Pi}{\bar{r}_{\hspace{-1pt},x}} \left( \frac{\bar{r}^2}{\bar{\lambda}}
        \right)_{\hspace{-3pt},x} \right]_{\hspace{-1pt},x}
        + \bar{r}_{,x} (\omega^2 W + Q) \right\rbrace \xi \nonumber \\
     && - \left\lbrace \frac{\bar{r}_{,x}}{\Pi}\left(\frac{\bar{\lambda}}{\bar{r}^2}\right)^2
        \left[ \left(\frac{\bar{r}^2}{\bar{\lambda}}\right)^2\frac{\Pi}{\bar{r}_{,x}}
        \right]_{,x} - \frac{\bar{C}^2_{,x}}{\bar{C}^2}\right\rbrace B\label{calcxi}\\
   0 &=& (\omega^2)_{,x}~\text{.}
 \end{eqnarray}

The last equation simply expresses the fact that the frequency
$\omega$ is constant. The purpose of introducing this equation becomes apparent when
we consider the boundary conditions. Spherical symmetry requires the
displacement $\xi$ to vanish at the origin. The auxiliary variable $B$
vanishes by definition at the stellar surface. The value of $B$ at the
origin determines the amplitude of the profile $\xi(x)$. Because
eigenmodes are only defined up to a constant factor, we are free to
choose for $B(0)$ an arbitrary finite value. We thus have a
two-point-boundary-value problem for the three variables
$\xi(x)$, $B(x)$ and $\omega(x)$ which we solve with a relaxation
algorithm. In particular, the solution for the constant $\omega(x)$
provides us with the frequency of the eigenmode. Solutions only exist
for specific values of the frequency and finding a given mode requires
a relatively careful initial guess.

\bibliography{paper.bib}

\begin{thebibliography}{65}
\expandafter\ifx\csname natexlab\endcsname\relax\def\natexlab#1{#1}\fi
\expandafter\ifx\csname bibnamefont\endcsname\relax
  \def\bibnamefont#1{#1}\fi
\expandafter\ifx\csname bibfnamefont\endcsname\relax
  \def\bibfnamefont#1{#1}\fi
\expandafter\ifx\csname citenamefont\endcsname\relax
  \def\citenamefont#1{#1}\fi
\expandafter\ifx\csname url\endcsname\relax
  \def\url#1{\texttt{#1}}\fi
\expandafter\ifx\csname urlprefix\endcsname\relax\def\urlprefix{URL }\fi
\providecommand{\bibinfo}[2]{#2}
\providecommand{\eprint}[2][]{\url{#2}}

\bibitem[{\citenamefont{Leavitt and Pickering}(1912)}]{Leavitt1912}
\bibinfo{author}{\bibfnamefont{H.~S.} \bibnamefont{Leavitt}} \bibnamefont{and}
  \bibinfo{author}{\bibfnamefont{E.~C.} \bibnamefont{Pickering}},
  \bibinfo{journal}{Harvard Coll. Obs. Circ.} \textbf{\bibinfo{volume}{173}},
  \bibinfo{pages}{1} (\bibinfo{year}{1912}).

\bibitem[{\citenamefont{Pickering}(1901)}]{Pickering1901}
\bibinfo{author}{\bibfnamefont{E.~C.} \bibnamefont{Pickering}},
  \bibinfo{journal}{Harvard Coll. Obs. Circ.} \textbf{\bibinfo{volume}{54}},
  \bibinfo{pages}{1} (\bibinfo{year}{1901}).

\bibitem[{\citenamefont{Alcock~{\em et al.}}(1995)}]{Alcock1995}
\bibinfo{author}{\bibfnamefont{C.}~\bibnamefont{Alcock~{\em et al.}}},
  \bibinfo{journal}{Astron. J.} \textbf{\bibinfo{volume}{109}},
  \bibinfo{pages}{1654} (\bibinfo{year}{1995}),
  \bibinfo{note}{astro-ph/9411061}.

\bibitem[{\citenamefont{Benedict~{\em at al.}}(2007)}]{Benedict2007}
\bibinfo{author}{\bibfnamefont{G.~F.} \bibnamefont{Benedict~{\em at al.}}},
  \bibinfo{journal}{Astron. J.} \textbf{\bibinfo{volume}{133}},
  \bibinfo{pages}{1810} (\bibinfo{year}{2007}),
  \bibinfo{note}{astro-ph/0612465}.

\bibitem[{\citenamefont{Olech and Moskalik}(2008)}]{Olech2008}
\bibinfo{author}{\bibfnamefont{A.}~\bibnamefont{Olech}} \bibnamefont{and}
  \bibinfo{author}{\bibfnamefont{P.}~\bibnamefont{Moskalik}}
  (\bibinfo{year}{2008}), \bibinfo{note}{arXiv:0812.4173 [astro-ph]}.

\bibitem[{\citenamefont{Birch}(2008)}]{Birch2008}
\bibinfo{author}{\bibfnamefont{A.~C.} \bibnamefont{Birch}},
  \bibinfo{journal}{Journal of Physics} \textbf{\bibinfo{volume}{118}},
  \bibinfo{pages}{012009} (\bibinfo{year}{2008}), \bibinfo{note}{proceedings of
  the Second HELAS International Conference}.

\bibitem[{\citenamefont{Chandrasekhar}(1964)}]{Chandrasekhar1964}
\bibinfo{author}{\bibfnamefont{S.}~\bibnamefont{Chandrasekhar}},
  \bibinfo{journal}{Phys. Rev. Letters} \textbf{\bibinfo{volume}{12}},
  \bibinfo{pages}{114} (\bibinfo{year}{1964}).

\bibitem[{\citenamefont{Sathyaprakash and Schutz}(2009)}]{Sathyaprakash2009}
\bibinfo{author}{\bibfnamefont{B.~S.} \bibnamefont{Sathyaprakash}}
  \bibnamefont{and} \bibinfo{author}{\bibfnamefont{B.~F.}
  \bibnamefont{Schutz}}, \bibinfo{journal}{Living Rev. Relativity}
  \textbf{\bibinfo{volume}{2009-12}} (\bibinfo{year}{2009}),
  \bibinfo{note}{[Article in Online Journal] cited on 6 Mar 2009
  http://relativity.livingreviews.org/Articles/lrr-2009-2/download/index.html}.

\bibitem[{\citenamefont{Chanmugam}(1977)}]{Chanmugam1977}
\bibinfo{author}{\bibfnamefont{G.}~\bibnamefont{Chanmugam}},
  \bibinfo{journal}{ApJ 217} \textbf{\bibinfo{volume}{217}},
  \bibinfo{pages}{799} (\bibinfo{year}{1977}).

\bibitem[{\citenamefont{Glass and Lindblom}(1983)}]{Glass1983}
\bibinfo{author}{\bibfnamefont{E.~N.} \bibnamefont{Glass}} \bibnamefont{and}
  \bibinfo{author}{\bibfnamefont{L.}~\bibnamefont{Lindblom}},
  \bibinfo{journal}{ApJ Suppl. Series} \textbf{\bibinfo{volume}{53}},
  \bibinfo{pages}{93} (\bibinfo{year}{1983}).

\bibitem[{\citenamefont{Vaeth and Chanmugam}(1992)}]{Vaeth1992}
\bibinfo{author}{\bibfnamefont{H.~M.} \bibnamefont{Vaeth}} \bibnamefont{and}
  \bibinfo{author}{\bibfnamefont{G.}~\bibnamefont{Chanmugam}},
  \bibinfo{journal}{Astronomy and Astrophysics 260}
  \textbf{\bibinfo{volume}{260}}, \bibinfo{pages}{250} (\bibinfo{year}{1992}).

\bibitem[{\citenamefont{Kokkotas and Ruoff}(2001)}]{Kokkotas2001}
\bibinfo{author}{\bibfnamefont{K.~D.} \bibnamefont{Kokkotas}} \bibnamefont{and}
  \bibinfo{author}{\bibfnamefont{J.}~\bibnamefont{Ruoff}},
  \bibinfo{journal}{Astron.Astrophys. 366} p. \bibinfo{pages}{565}
  (\bibinfo{year}{2001}).

\bibitem[{\citenamefont{Koll{\'a}th et~al.}(1998)\citenamefont{Koll{\'a}th,
  Beaulieu, Buchler, and Yecko}}]{Kollath1998}
\bibinfo{author}{\bibfnamefont{Z.}~\bibnamefont{Koll{\'a}th}},
  \bibinfo{author}{\bibfnamefont{J.~P.} \bibnamefont{Beaulieu}},
  \bibinfo{author}{\bibfnamefont{J.~R.} \bibnamefont{Buchler}},
  \bibnamefont{and} \bibinfo{author}{\bibfnamefont{P.}~\bibnamefont{Yecko}},
  \bibinfo{journal}{Astrophys. J.} \textbf{\bibinfo{volume}{502}},
  \bibinfo{pages}{L55} (\bibinfo{year}{1998}),
  \bibinfo{note}{astro-ph/9804111}.

\bibitem[{\citenamefont{{Schenk} et~al.}(2002)\citenamefont{{Schenk}, {Arras},
  {Flanagan}, {Teukolsky}, and {Wasserman}}}]{Schenk2002}
\bibinfo{author}{\bibfnamefont{A.~K.} \bibnamefont{{Schenk}}},
  \bibinfo{author}{\bibfnamefont{P.}~\bibnamefont{{Arras}}},
  \bibinfo{author}{\bibfnamefont{{\'E}.~{\'E}.} \bibnamefont{{Flanagan}}},
  \bibinfo{author}{\bibfnamefont{S.~A.} \bibnamefont{{Teukolsky}}},
  \bibnamefont{and}
  \bibinfo{author}{\bibfnamefont{I.}~\bibnamefont{{Wasserman}}},
  \bibinfo{journal}{Phys.Rev. D 65}  (\bibinfo{year}{2002}).

\bibitem[{\citenamefont{{Arras} et~al.}(2003)\citenamefont{{Arras}, {Flanagan},
  {Morsink}, {Schenk}, {Teukolsky}, and {Wasserman}}}]{Arras2003}
\bibinfo{author}{\bibfnamefont{P.}~\bibnamefont{{Arras}}},
  \bibinfo{author}{\bibfnamefont{E.~E.} \bibnamefont{{Flanagan}}},
  \bibinfo{author}{\bibfnamefont{S.~M.} \bibnamefont{{Morsink}}},
  \bibinfo{author}{\bibfnamefont{A.~K.} \bibnamefont{{Schenk}}},
  \bibinfo{author}{\bibfnamefont{S.~A.} \bibnamefont{{Teukolsky}}},
  \bibnamefont{and}
  \bibinfo{author}{\bibfnamefont{I.}~\bibnamefont{{Wasserman}}},
  \bibinfo{journal}{ApJ} \textbf{\bibinfo{volume}{591}}, \bibinfo{pages}{1129}
  (\bibinfo{year}{2003}).

\bibitem[{\citenamefont{Brink et~al.}(2004)\citenamefont{Brink, Teukolsky, and
  Wasserman}}]{Brink2004}
\bibinfo{author}{\bibfnamefont{J.}~\bibnamefont{Brink}},
  \bibinfo{author}{\bibfnamefont{S.~A.} \bibnamefont{Teukolsky}},
  \bibnamefont{and}
  \bibinfo{author}{\bibfnamefont{I.}~\bibnamefont{Wasserman}},
  \bibinfo{journal}{Phys. Rev.} \textbf{\bibinfo{volume}{D70}},
  \bibinfo{pages}{124017} (\bibinfo{year}{2004}).

\bibitem[{\citenamefont{Brink et~al.}(2005)\citenamefont{Brink, Teukolsky, and
  Wasserman}}]{Brink2004b}
\bibinfo{author}{\bibfnamefont{J.}~\bibnamefont{Brink}},
  \bibinfo{author}{\bibfnamefont{S.~A.} \bibnamefont{Teukolsky}},
  \bibnamefont{and}
  \bibinfo{author}{\bibfnamefont{I.}~\bibnamefont{Wasserman}},
  \bibinfo{journal}{Phys. Rev.} \textbf{\bibinfo{volume}{D71}},
  \bibinfo{pages}{064029} (\bibinfo{year}{2005}).

\bibitem[{\citenamefont{{Lin} and {Suen}}(2006)}]{Lin2006}
\bibinfo{author}{\bibfnamefont{L.}~\bibnamefont{{Lin}}} \bibnamefont{and}
  \bibinfo{author}{\bibfnamefont{W.}~\bibnamefont{{Suen}}},
  \bibinfo{journal}{Mon. Not. Roy. Astron. Soc.}
  \textbf{\bibinfo{volume}{370}}, \bibinfo{pages}{1295} (\bibinfo{year}{2006}).

\bibitem[{\citenamefont{{Bondarescu} et~al.}(2008)\citenamefont{{Bondarescu},
  {Teukolsky}, and {Wasserman}}}]{Bondarescu2008}
\bibinfo{author}{\bibfnamefont{R.}~\bibnamefont{{Bondarescu}}},
  \bibinfo{author}{\bibfnamefont{S.~A.} \bibnamefont{{Teukolsky}}},
  \bibnamefont{and}
  \bibinfo{author}{\bibfnamefont{I.}~\bibnamefont{{Wasserman}}},
  \bibinfo{journal}{ArXiv e-prints}  (\bibinfo{year}{2008}),
  \eprint{0809.3448}.

\bibitem[{\citenamefont{Dziembowski}(1982)}]{Dziembowski1982}
\bibinfo{author}{\bibfnamefont{W.}~\bibnamefont{Dziembowski}},
  \bibinfo{journal}{Acta Astronomica} \textbf{\bibinfo{volume}{32}},
  \bibinfo{pages}{147} (\bibinfo{year}{1982}).

\bibitem[{\citenamefont{Kumar and Goldreich}(1989)}]{Kumar1989}
\bibinfo{author}{\bibfnamefont{P.}~\bibnamefont{Kumar}} \bibnamefont{and}
  \bibinfo{author}{\bibfnamefont{P.}~\bibnamefont{Goldreich}},
  \bibinfo{journal}{ApJ} \textbf{\bibinfo{volume}{342}}, \bibinfo{pages}{558}
  (\bibinfo{year}{1989}).

\bibitem[{\citenamefont{Van~Hoolst}(1996)}]{VanHoolst1996}
\bibinfo{author}{\bibfnamefont{T.}~\bibnamefont{Van~Hoolst}},
  \bibinfo{journal}{Astronomy and Astrophysics} \textbf{\bibinfo{volume}{308}},
  \bibinfo{pages}{66} (\bibinfo{year}{1996}).

\bibitem[{\citenamefont{Passamonti et~al.}(2006)\citenamefont{Passamonti,
  Bruni, Gualtieri, Nagar, and Sopuerta}}]{Passamonti2006}
\bibinfo{author}{\bibfnamefont{A.}~\bibnamefont{Passamonti}},
  \bibinfo{author}{\bibfnamefont{M.}~\bibnamefont{Bruni}},
  \bibinfo{author}{\bibfnamefont{L.}~\bibnamefont{Gualtieri}},
  \bibinfo{author}{\bibfnamefont{A.}~\bibnamefont{Nagar}}, \bibnamefont{and}
  \bibinfo{author}{\bibfnamefont{C.~F.} \bibnamefont{Sopuerta}},
  \bibinfo{journal}{Phys. Rev. D} \textbf{\bibinfo{volume}{73}},
  \bibinfo{pages}{084010} (\bibinfo{year}{2006}),
  \bibinfo{note}{gr-qc/0601001}.

\bibitem[{\citenamefont{Passamonti et~al.}(2007)\citenamefont{Passamonti,
  Stergioulas, and Nagar}}]{Passamonti2007}
\bibinfo{author}{\bibfnamefont{A.}~\bibnamefont{Passamonti}},
  \bibinfo{author}{\bibfnamefont{N.}~\bibnamefont{Stergioulas}},
  \bibnamefont{and} \bibinfo{author}{\bibfnamefont{A.}~\bibnamefont{Nagar}},
  \bibinfo{journal}{Phys. Rev.} \textbf{\bibinfo{volume}{D75}},
  \bibinfo{pages}{084038} (\bibinfo{year}{2007}).

\bibitem[{\citenamefont{Brizuela et~al.}(2006)\citenamefont{Brizuela,
  Mart{\'i}n-Garc{\'i}a, and Mena-Marugan}}]{Brizuela2006}
\bibinfo{author}{\bibfnamefont{D.}~\bibnamefont{Brizuela}},
  \bibinfo{author}{\bibfnamefont{J.~M.} \bibnamefont{Mart{\'i}n-Garc{\'i}a}},
  \bibnamefont{and} \bibinfo{author}{\bibfnamefont{G.~A.}
  \bibnamefont{Mena-Marugan}}, \bibinfo{journal}{Phys. Rev. D}
  \textbf{\bibinfo{volume}{74}}, \bibinfo{pages}{044039}
  (\bibinfo{year}{2006}), \bibinfo{note}{gr-qc/0607025}.

\bibitem[{\citenamefont{Brizuela et~al.}(2007)\citenamefont{Brizuela,
  Mart{\'i}n-Garc{\'i}a, and Mena-Marugan}}]{Brizuela2007}
\bibinfo{author}{\bibfnamefont{D.}~\bibnamefont{Brizuela}},
  \bibinfo{author}{\bibfnamefont{J.~M.} \bibnamefont{Mart{\'i}n-Garc{\'i}a}},
  \bibnamefont{and} \bibinfo{author}{\bibfnamefont{G.~A.}
  \bibnamefont{Mena-Marugan}}, \bibinfo{journal}{Phys. Rev. D}
  \textbf{\bibinfo{volume}{76}}, \bibinfo{pages}{024004}
  (\bibinfo{year}{2007}), \bibinfo{note}{gr-qc/0703069}.

\bibitem[{\citenamefont{Brizuela et~al.}(2009)\citenamefont{Brizuela,
  Mart{\'i}n-Garc{\'i}a, and Tiglio}}]{Brizuela2009a}
\bibinfo{author}{\bibfnamefont{D.}~\bibnamefont{Brizuela}},
  \bibinfo{author}{\bibfnamefont{J.~M.} \bibnamefont{Mart{\'i}n-Garc{\'i}a}},
  \bibnamefont{and} \bibinfo{author}{\bibfnamefont{M.}~\bibnamefont{Tiglio}}
  (\bibinfo{year}{2009}), \bibinfo{note}{0903.1134}.

\bibitem[{\citenamefont{Brizuela et~al.}()\citenamefont{Brizuela,
  Mart{\'i}n-Garc{\'i}a, Sperhake, and Kokkotas}}]{Brizuela2009}
\bibinfo{author}{\bibfnamefont{D.}~\bibnamefont{Brizuela}},
  \bibinfo{author}{\bibfnamefont{J.~M.} \bibnamefont{Mart{\'i}n-Garc{\'i}a}},
  \bibinfo{author}{\bibfnamefont{U.}~\bibnamefont{Sperhake}}, \bibnamefont{and}
  \bibinfo{author}{\bibfnamefont{K.}~\bibnamefont{Kokkotas}},
  \emph{\bibinfo{title}{High-order perturbations pf a spherical collapsing star
  (forthcoming)}}.

\bibitem[{\citenamefont{Berti et~al.}(2006)\citenamefont{Berti, Cardoso, and
  Will}}]{Berti2006}
\bibinfo{author}{\bibfnamefont{E.}~\bibnamefont{Berti}},
  \bibinfo{author}{\bibfnamefont{V.}~\bibnamefont{Cardoso}}, \bibnamefont{and}
  \bibinfo{author}{\bibfnamefont{C.~M.} \bibnamefont{Will}},
  \bibinfo{journal}{Phys. Rev. D} \textbf{\bibinfo{volume}{73}},
  \bibinfo{pages}{064030} (\bibinfo{year}{2006}),
  \bibinfo{note}{gr-qc/0512160}.

\bibitem[{\citenamefont{Ferrari and Gualtieri}(2008)}]{Ferrari2007}
\bibinfo{author}{\bibfnamefont{V.}~\bibnamefont{Ferrari}} \bibnamefont{and}
  \bibinfo{author}{\bibfnamefont{L.}~\bibnamefont{Gualtieri}},
  \bibinfo{journal}{Gen. Rel. Grav.} \textbf{\bibinfo{volume}{40}},
  \bibinfo{pages}{945} (\bibinfo{year}{2008}), \bibinfo{note}{arXiv:0709.0657
  [gr-qc]}.

\bibitem[{\citenamefont{Leaver}(1986)}]{Leaver1986}
\bibinfo{author}{\bibfnamefont{E.~W.} \bibnamefont{Leaver}},
  \bibinfo{journal}{Phys. Rev. D} \textbf{\bibinfo{volume}{34}},
  \bibinfo{pages}{384} (\bibinfo{year}{1986}).

\bibitem[{\citenamefont{Kokkotas}(1999)}]{Kokkotas1999}
\bibinfo{author}{\bibfnamefont{K.}~\bibnamefont{Kokkotas}},
  \bibinfo{journal}{Living Rev. Relativity} \textbf{\bibinfo{volume}{1999-2}}
  (\bibinfo{year}{1999}), \bibinfo{note}{[Article in Online Journal] cited on
  13 Apr 2008,
  http://relativity.livingreviews.org/Articles/lrr-1999-2/download/index.html}.

\bibitem[{\citenamefont{Berti et~al.}(2007{\natexlab{a}})\citenamefont{Berti,
  Cardoso, Gonz{\'a}lez, Sperhake, Hannam, Husa, and Br{\"u}gmann}}]{Berti2007}
\bibinfo{author}{\bibfnamefont{E.}~\bibnamefont{Berti}},
  \bibinfo{author}{\bibfnamefont{V.}~\bibnamefont{Cardoso}},
  \bibinfo{author}{\bibfnamefont{J.~A.} \bibnamefont{Gonz{\'a}lez}},
  \bibinfo{author}{\bibfnamefont{U.}~\bibnamefont{Sperhake}},
  \bibinfo{author}{\bibfnamefont{M.~D.} \bibnamefont{Hannam}},
  \bibinfo{author}{\bibfnamefont{S.}~\bibnamefont{Husa}}, \bibnamefont{and}
  \bibinfo{author}{\bibfnamefont{B.}~\bibnamefont{Br{\"u}gmann}},
  \bibinfo{journal}{Phys. Rev. D} \textbf{\bibinfo{volume}{76}},
  \bibinfo{pages}{064034} (\bibinfo{year}{2007}{\natexlab{a}}),
  \bibinfo{note}{gr-qc/0703053}.

\bibitem[{\citenamefont{Nakano and Ioka}(2007)}]{Nakano2007}
\bibinfo{author}{\bibfnamefont{H.}~\bibnamefont{Nakano}} \bibnamefont{and}
  \bibinfo{author}{\bibfnamefont{K.}~\bibnamefont{Ioka}},
  \bibinfo{journal}{Phys. Rev. D} \textbf{\bibinfo{volume}{76}},
  \bibinfo{pages}{084007} (\bibinfo{year}{2007}),
  \bibinfo{note}{arXiv:0708.0450 [gr-qc]}.

\bibitem[{\citenamefont{Buonanno et~al.}(2007)\citenamefont{Buonanno, Cook, and
  Pretorius}}]{Buonanno2006}
\bibinfo{author}{\bibfnamefont{A.}~\bibnamefont{Buonanno}},
  \bibinfo{author}{\bibfnamefont{G.~B.} \bibnamefont{Cook}}, \bibnamefont{and}
  \bibinfo{author}{\bibfnamefont{F.}~\bibnamefont{Pretorius}},
  \bibinfo{journal}{Phys. Rev. D} \textbf{\bibinfo{volume}{75}},
  \bibinfo{pages}{124018} (\bibinfo{year}{2007}),
  \bibinfo{note}{gr-qc/0610122}.

\bibitem[{\citenamefont{Berti et~al.}(2007{\natexlab{b}})\citenamefont{Berti,
  Cardoso, Gonz{\'a}lez, and Sperhake}}]{Berti2007b}
\bibinfo{author}{\bibfnamefont{E.}~\bibnamefont{Berti}},
  \bibinfo{author}{\bibfnamefont{V.}~\bibnamefont{Cardoso}},
  \bibinfo{author}{\bibfnamefont{J.~A.} \bibnamefont{Gonz{\'a}lez}},
  \bibnamefont{and} \bibinfo{author}{\bibfnamefont{U.}~\bibnamefont{Sperhake}},
  \bibinfo{journal}{Phys. Rev. D} \textbf{\bibinfo{volume}{75}},
  \bibinfo{pages}{124017} (\bibinfo{year}{2007}{\natexlab{b}}),
  \bibinfo{note}{gr-qc/0701086}.

\bibitem[{\citenamefont{Baker et~al.}(2008)\citenamefont{Baker, Boggs,
  Centrella, Kelly, McWilliams, and van Meter}}]{Baker2008a}
\bibinfo{author}{\bibfnamefont{J.~G.} \bibnamefont{Baker}},
  \bibinfo{author}{\bibfnamefont{W.~D.} \bibnamefont{Boggs}},
  \bibinfo{author}{\bibfnamefont{J.}~\bibnamefont{Centrella}},
  \bibinfo{author}{\bibfnamefont{B.~J.} \bibnamefont{Kelly}},
  \bibinfo{author}{\bibfnamefont{S.~T.} \bibnamefont{McWilliams}},
  \bibnamefont{and} \bibinfo{author}{\bibfnamefont{J.~R.} \bibnamefont{van
  Meter}}, \bibinfo{journal}{Phys. Rev. D} \textbf{\bibinfo{volume}{78}},
  \bibinfo{pages}{044046} (\bibinfo{year}{2008}),
  \bibinfo{note}{arXiv:0805.1428 [gr-qc]}.

\bibitem[{\citenamefont{Papadopoulos and Sopuerta}(2002)}]{Papadopoulos2001}
\bibinfo{author}{\bibfnamefont{P.}~\bibnamefont{Papadopoulos}}
  \bibnamefont{and} \bibinfo{author}{\bibfnamefont{C.~F.}
  \bibnamefont{Sopuerta}}, \bibinfo{journal}{Phys. Rev. D}
  \textbf{\bibinfo{volume}{65}}, \bibinfo{pages}{044008}
  (\bibinfo{year}{2002}), \bibinfo{note}{gr-qc/0107051}.

\bibitem[{\citenamefont{Lindblom
  et~al.}(2001{\natexlab{a}})\citenamefont{Lindblom, Tohline, and
  Vallisneri}}]{Lindblom2000}
\bibinfo{author}{\bibfnamefont{L.}~\bibnamefont{Lindblom}},
  \bibinfo{author}{\bibfnamefont{J.~E.} \bibnamefont{Tohline}},
  \bibnamefont{and}
  \bibinfo{author}{\bibfnamefont{M.}~\bibnamefont{Vallisneri}},
  \bibinfo{journal}{Phys. Rev. Lett.} \textbf{\bibinfo{volume}{86}},
  \bibinfo{pages}{1152} (\bibinfo{year}{2001}{\natexlab{a}}),
  \bibinfo{note}{astro-ph/0010653}.

\bibitem[{\citenamefont{Lindblom
  et~al.}(2001{\natexlab{b}})\citenamefont{Lindblom, Tohline, and
  Vallisneri}}]{Lindblom2001}
\bibinfo{author}{\bibfnamefont{L.}~\bibnamefont{Lindblom}},
  \bibinfo{author}{\bibfnamefont{J.~E.} \bibnamefont{Tohline}},
  \bibnamefont{and}
  \bibinfo{author}{\bibfnamefont{M.}~\bibnamefont{Vallisneri}},
  \bibinfo{journal}{Phys. Rev. Lett.} \textbf{\bibinfo{volume}{86}},
  \bibinfo{pages}{1152} (\bibinfo{year}{2001}{\natexlab{b}}),
  \eprint{arXiv:astro-ph/0010653}.

\bibitem[{\citenamefont{{Stergioulas} and {Font}}(2001)}]{Stergioulas2001}
\bibinfo{author}{\bibfnamefont{N.}~\bibnamefont{{Stergioulas}}}
  \bibnamefont{and} \bibinfo{author}{\bibfnamefont{J.~A.}
  \bibnamefont{{Font}}}, \bibinfo{journal}{Phys. Rev. Lett.}
  \textbf{\bibinfo{volume}{86}}, \bibinfo{pages}{1148} (\bibinfo{year}{2001}).

\bibitem[{\citenamefont{Font et~al.}(2001)\citenamefont{Font, Dimmelmeier,
  Gupta, and Stergioulas}}]{Font2001}
\bibinfo{author}{\bibfnamefont{J.}~\bibnamefont{Font}},
  \bibinfo{author}{\bibfnamefont{H.}~\bibnamefont{Dimmelmeier}},
  \bibinfo{author}{\bibfnamefont{A.}~\bibnamefont{Gupta}}, \bibnamefont{and}
  \bibinfo{author}{\bibfnamefont{N.}~\bibnamefont{Stergioulas}},
  \bibinfo{journal}{Mon. Not. Roy. Astron. Soc.}
  \textbf{\bibinfo{volume}{325}}, \bibinfo{pages}{1463} (\bibinfo{year}{2001}).

\bibitem[{\citenamefont{Stergioulas et~al.}(2004)\citenamefont{Stergioulas,
  Apostolatos, and Font}}]{Stergioulas2004}
\bibinfo{author}{\bibfnamefont{N.}~\bibnamefont{Stergioulas}},
  \bibinfo{author}{\bibfnamefont{T.~A.} \bibnamefont{Apostolatos}},
  \bibnamefont{and} \bibinfo{author}{\bibfnamefont{J.~A.} \bibnamefont{Font}},
  \bibinfo{journal}{Mon. Not. Roy. Astron. Soc.}
  \textbf{\bibinfo{volume}{352}}, \bibinfo{pages}{1089} (\bibinfo{year}{2004}).

\bibitem[{\citenamefont{Dimmelmeier et~al.}(2006)\citenamefont{Dimmelmeier,
  Stergioulas, and Font}}]{Dimmelmeier2006}
\bibinfo{author}{\bibfnamefont{H.}~\bibnamefont{Dimmelmeier}},
  \bibinfo{author}{\bibfnamefont{N.}~\bibnamefont{Stergioulas}},
  \bibnamefont{and} \bibinfo{author}{\bibfnamefont{J.~A.} \bibnamefont{Font}},
  \bibinfo{journal}{Mon. Not. Roy. Astron. Soc.}
  \textbf{\bibinfo{volume}{368}}, \bibinfo{pages}{1609} (\bibinfo{year}{2006}).

\bibitem[{\citenamefont{Font et~al.}(2002)\citenamefont{Font, Goodale, Iyer,
  Miller, Rezzolla, Seidel, Stergioulas, Suen, and Tobias}}]{Font2002}
\bibinfo{author}{\bibfnamefont{J.~A.} \bibnamefont{Font}},
  \bibinfo{author}{\bibfnamefont{T.}~\bibnamefont{Goodale}},
  \bibinfo{author}{\bibfnamefont{S.}~\bibnamefont{Iyer}},
  \bibinfo{author}{\bibfnamefont{M.}~\bibnamefont{Miller}},
  \bibinfo{author}{\bibfnamefont{L.}~\bibnamefont{Rezzolla}},
  \bibinfo{author}{\bibfnamefont{E.}~\bibnamefont{Seidel}},
  \bibinfo{author}{\bibfnamefont{N.}~\bibnamefont{Stergioulas}},
  \bibinfo{author}{\bibfnamefont{W.~M.} \bibnamefont{Suen}}, \bibnamefont{and}
  \bibinfo{author}{\bibfnamefont{M.}~\bibnamefont{Tobias}},
  \bibinfo{journal}{Phys. Rev. D} \textbf{\bibinfo{volume}{65}},
  \bibinfo{pages}{084024} (\bibinfo{year}{2002}).

\bibitem[{\citenamefont{Baiotti~{\em et al.}}(2005)}]{Baiotti2004}
\bibinfo{author}{\bibfnamefont{L.}~\bibnamefont{Baiotti~{\em et al.}}},
  \bibinfo{journal}{Phys. Rev. D} \textbf{\bibinfo{volume}{D71}},
  \bibinfo{pages}{024035} (\bibinfo{year}{2005}),
  \bibinfo{note}{gr-qc/0403029}.

\bibitem[{\citenamefont{Marronetti et~al.}(2004)\citenamefont{Marronetti, Duez,
  Shapiro, and Baumgarte}}]{Marronetti2004}
\bibinfo{author}{\bibfnamefont{P.}~\bibnamefont{Marronetti}},
  \bibinfo{author}{\bibfnamefont{M.~D.} \bibnamefont{Duez}},
  \bibinfo{author}{\bibfnamefont{S.~L.} \bibnamefont{Shapiro}},
  \bibnamefont{and} \bibinfo{author}{\bibfnamefont{T.~W.}
  \bibnamefont{Baumgarte}}, \bibinfo{journal}{Phys. Rev. Lett.}
  \textbf{\bibinfo{volume}{92}}, \bibinfo{pages}{141101}
  (\bibinfo{year}{2004}), \bibinfo{note}{gr-qc/0312036}.

\bibitem[{\citenamefont{Miller et~al.}(2004)\citenamefont{Miller, Gressman, and
  Suen}}]{Miller2004}
\bibinfo{author}{\bibfnamefont{M.}~\bibnamefont{Miller}},
  \bibinfo{author}{\bibfnamefont{P.}~\bibnamefont{Gressman}}, \bibnamefont{and}
  \bibinfo{author}{\bibfnamefont{W.-M.} \bibnamefont{Suen}},
  \bibinfo{journal}{Phys. Rev. D} \textbf{\bibinfo{volume}{69}},
  \bibinfo{pages}{064026} (\bibinfo{year}{2004}),
  \bibinfo{note}{gr-qc/0312030}.

\bibitem[{\citenamefont{Shibata and Taniguchi}(2006)}]{Shibata2006}
\bibinfo{author}{\bibfnamefont{M.}~\bibnamefont{Shibata}} \bibnamefont{and}
  \bibinfo{author}{\bibfnamefont{K.}~\bibnamefont{Taniguchi}},
  \bibinfo{journal}{Phys. Rev. D} \textbf{\bibinfo{volume}{73}},
  \bibinfo{pages}{064027} (\bibinfo{year}{2006}),
  \bibinfo{note}{astro-ph/0603145}.

\bibitem[{\citenamefont{Duez et~al.}(2006)\citenamefont{Duez, Liu, Shapiro,
  Shibata, and Stephens}}]{Duez2006}
\bibinfo{author}{\bibfnamefont{M.~D.} \bibnamefont{Duez}},
  \bibinfo{author}{\bibfnamefont{Y.~T.} \bibnamefont{Liu}},
  \bibinfo{author}{\bibfnamefont{S.~L.} \bibnamefont{Shapiro}},
  \bibinfo{author}{\bibfnamefont{M.}~\bibnamefont{Shibata}}, \bibnamefont{and}
  \bibinfo{author}{\bibfnamefont{B.~C.} \bibnamefont{Stephens}},
  \bibinfo{journal}{Phys. Rev. D} \textbf{\bibinfo{volume}{73}},
  \bibinfo{pages}{104015} (\bibinfo{year}{2006}),
  \bibinfo{note}{gr-qc/0605331}.

\bibitem[{\citenamefont{Anderson et~al.}(2008)\citenamefont{Anderson,
  Hirschmann, Lehner, Liebling, Motl, Neilsen, Palenzuela, and
  Tohline}}]{Anderson2007}
\bibinfo{author}{\bibfnamefont{M.}~\bibnamefont{Anderson}},
  \bibinfo{author}{\bibfnamefont{E.~W.} \bibnamefont{Hirschmann}},
  \bibinfo{author}{\bibfnamefont{L.}~\bibnamefont{Lehner}},
  \bibinfo{author}{\bibfnamefont{S.~L.} \bibnamefont{Liebling}},
  \bibinfo{author}{\bibfnamefont{P.~M.} \bibnamefont{Motl}},
  \bibinfo{author}{\bibfnamefont{D.}~\bibnamefont{Neilsen}},
  \bibinfo{author}{\bibfnamefont{C.}~\bibnamefont{Palenzuela}},
  \bibnamefont{and} \bibinfo{author}{\bibfnamefont{J.~E.}
  \bibnamefont{Tohline}}, \bibinfo{journal}{Phys. Rev. D}
  \textbf{\bibinfo{volume}{77}}, \bibinfo{pages}{024006}
  (\bibinfo{year}{2008}), \bibinfo{note}{arXiv:0708:2720 [gr-qc]}.

\bibitem[{\citenamefont{Duez et~al.}(2008)\citenamefont{Duez, Foucart, Kidder,
  Pfeiffer, Scheel, and Teukolsky}}]{Duez2008}
\bibinfo{author}{\bibfnamefont{M.~D.} \bibnamefont{Duez}},
  \bibinfo{author}{\bibfnamefont{F.}~\bibnamefont{Foucart}},
  \bibinfo{author}{\bibfnamefont{L.~E.} \bibnamefont{Kidder}},
  \bibinfo{author}{\bibfnamefont{H.~P.} \bibnamefont{Pfeiffer}},
  \bibinfo{author}{\bibfnamefont{M.~A.} \bibnamefont{Scheel}},
  \bibnamefont{and} \bibinfo{author}{\bibfnamefont{S.~A.}
  \bibnamefont{Teukolsky}}, \bibinfo{journal}{Phys. Rev. D}
  \textbf{\bibinfo{volume}{78}}, \bibinfo{pages}{104015}
  (\bibinfo{year}{2008}), \bibinfo{note}{arXiv:0809.0002 [gr-qc]}.

\bibitem[{\citenamefont{Baiotti et~al.}(2008)\citenamefont{Baiotti, Giacomazzo,
  and Rezzolla}}]{Baiotti2008}
\bibinfo{author}{\bibfnamefont{L.}~\bibnamefont{Baiotti}},
  \bibinfo{author}{\bibfnamefont{B.}~\bibnamefont{Giacomazzo}},
  \bibnamefont{and} \bibinfo{author}{\bibfnamefont{L.}~\bibnamefont{Rezzolla}},
  \bibinfo{journal}{Phys. Rev. D} \textbf{\bibinfo{volume}{78}},
  \bibinfo{pages}{084033} (\bibinfo{year}{2008}),
  \bibinfo{note}{arXiv:0804.0594 [gr-qc]}.

\bibitem[{\citenamefont{Sperhake et~al.}(2001)\citenamefont{Sperhake,
  Papadopoulos, and Andersson}}]{Sperhake2001paper}
\bibinfo{author}{\bibfnamefont{U.}~\bibnamefont{Sperhake}},
  \bibinfo{author}{\bibfnamefont{P.}~\bibnamefont{Papadopoulos}},
  \bibnamefont{and}
  \bibinfo{author}{\bibfnamefont{N.}~\bibnamefont{Andersson}},
  \emph{\bibinfo{title}{Non-linear radial oscillations of neutron stars:
  Mode-coupling results}} (\bibinfo{year}{2001}),
  \eprint{arXiv:astro-ph/0110487}.

\bibitem[{\citenamefont{May and White}(1966)}]{MayWhite1966}
\bibinfo{author}{\bibfnamefont{M.}~\bibnamefont{May}} \bibnamefont{and}
  \bibinfo{author}{\bibfnamefont{R.}~\bibnamefont{White}},
  \bibinfo{journal}{Phys.Rev. 141 Nr. 4} pp. \bibinfo{pages}{1232--1241}
  (\bibinfo{year}{1966}).

\bibitem[{\citenamefont{Misner and Sharp}(1964)}]{MisnerSharp1964}
\bibinfo{author}{\bibfnamefont{C.}~\bibnamefont{Misner}} \bibnamefont{and}
  \bibinfo{author}{\bibfnamefont{D.}~\bibnamefont{Sharp}},
  \bibinfo{journal}{Phys.Rev. 136 Nr. 2B} pp. \bibinfo{pages}{B571--B576}
  (\bibinfo{year}{1964}).

\bibitem[{\citenamefont{Seidel}(1990)}]{Seidel1990}
\bibinfo{author}{\bibfnamefont{E.}~\bibnamefont{Seidel}},
  \bibinfo{journal}{Phys. Rev. D} \textbf{\bibinfo{volume}{42}},
  \bibinfo{pages}{1884} (\bibinfo{year}{1990}).

\bibitem[{\citenamefont{Tolman}(1939)}]{Tolman1939}
\bibinfo{author}{\bibfnamefont{R.}~\bibnamefont{Tolman}},
  \bibinfo{journal}{Phys.Rev. 55} pp. \bibinfo{pages}{364--373}
  (\bibinfo{year}{1939}).

\bibitem[{\citenamefont{Oppenheimer and Volkoff}(1939)}]{OpVol1939}
\bibinfo{author}{\bibfnamefont{J.}~\bibnamefont{Oppenheimer}} \bibnamefont{and}
  \bibinfo{author}{\bibfnamefont{G.}~\bibnamefont{Volkoff}},
  \bibinfo{journal}{Phys.Rev. 55} pp. \bibinfo{pages}{374--381}
  (\bibinfo{year}{1939}).

\bibitem[{\citenamefont{Misner et~al.}(1973)\citenamefont{Misner, Thorne, and
  Wheeler}}]{Misner1973}
\bibinfo{author}{\bibfnamefont{C.}~\bibnamefont{Misner}},
  \bibinfo{author}{\bibfnamefont{K.}~\bibnamefont{Thorne}}, \bibnamefont{and}
  \bibinfo{author}{\bibfnamefont{J.}~\bibnamefont{Wheeler}},
  \emph{\bibinfo{title}{Gravitation}} (\bibinfo{publisher}{W.H. Freeman},
  \bibinfo{year}{1973}).

\bibitem[{\citenamefont{Press et~al.}(1992)\citenamefont{Press, Teukosky,
  Vetterling, and Flannery}}]{Press1992}
\bibinfo{author}{\bibfnamefont{W.}~\bibnamefont{Press}},
  \bibinfo{author}{\bibfnamefont{S.}~\bibnamefont{Teukosky}},
  \bibinfo{author}{\bibfnamefont{W.}~\bibnamefont{Vetterling}},
  \bibnamefont{and} \bibinfo{author}{\bibfnamefont{B.}~\bibnamefont{Flannery}},
  \emph{\bibinfo{title}{Numerical Recipes in C}} (\bibinfo{publisher}{Cambridge
  University Press, Cambridge}, \bibinfo{year}{1992}), \bibinfo{edition}{2nd}
  ed.

\bibitem[{\citenamefont{Sperhake}(2001)}]{Sperhake2001}
\bibinfo{author}{\bibfnamefont{U.}~\bibnamefont{Sperhake}}, Ph.D. thesis,
  \bibinfo{school}{University of Southampton} (\bibinfo{year}{2001}).

\bibitem[{\citenamefont{Ruoff}(2000)}]{Ruoff2000}
\bibinfo{author}{\bibfnamefont{J.}~\bibnamefont{Ruoff}}, Ph.D. thesis,
  \bibinfo{school}{Fakult\"{a}t f\"{u}r Physik der
  Eberhard-Karls-Universit\"{a}t T\"{u}bingen} (\bibinfo{year}{2000}).

\bibitem[{\citenamefont{Gundlach and Please}(2009)}]{Gundlach2009}
\bibinfo{author}{\bibfnamefont{C.}~\bibnamefont{Gundlach}} \bibnamefont{and}
  \bibinfo{author}{\bibfnamefont{C.}~\bibnamefont{Please}},
  \bibinfo{journal}{Phys. Rev. D} \textbf{\bibinfo{volume}{79}},
  \bibinfo{pages}{067501} (\bibinfo{year}{2009}).

\bibitem[{\citenamefont{Sethian}(1999)}]{Sethian1999}
\bibinfo{author}{\bibfnamefont{J.~A.} \bibnamefont{Sethian}},
  \emph{\bibinfo{title}{Level {S}et {M}ethods and {F}ast {M}arching {M}ethods}}
  (\bibinfo{publisher}{Cambridge University Press}, \bibinfo{year}{1999}),
  \bibinfo{note}{second edition}.

\end{thebibliography}

\end{document}